\documentclass[12pt,a4paper,fleqn]{article}

\usepackage[textheight=23cm,textwidth=15cm]{geometry}
\usepackage{amsmath}
\usepackage{graphicx}
\usepackage[american]{babel}
\usepackage{here}
\usepackage[articletitle=true, super=false]{achemso}
\setcitestyle{square}

\usepackage{hyperref} 
\hypersetup{colorlinks,linkcolor=black,urlcolor=blue, citecolor=black}
\usepackage{upgreek} 
\usepackage[version=4]{mhchem} 

\begin{document}

\begin{center}

{\LARGE\bf
Autonomous Reaction Network Exploration in Homogeneous and Heterogeneous Catalysis
}

\vspace{1cm}

{\large
Miguel Steiner\footnote{ORCID: 0000-0002-7634-7268} and
Markus Reiher\footnote{Corresponding author; e-mail: markus.reiher@phys.chem.ethz.ch; ORCID: 0000-0002-9508-1565}
}\\[4ex]

Laboratory of Physical Chemistry, ETH Zurich, \\
Vladimir-Prelog-Weg 2, 8093 Zurich, Switzerland

November 14, 2021

\vspace{.43cm}

\textbf{Abstract}
\end{center}
\vspace*{-.41cm}
{\small
Autonomous computations that rely on automated reaction network
elucidation algorithms may pave the way to make computational
catalysis on a par with experimental research in the field.
Several advantages of this approach are key to catalysis: 
(i) Automation allows one to consider orders of magnitude more 
structures in a systematic and open-ended fashion than what would be accessible by manual inspection. 
Eventually, full resolution in terms of structural varieties and conformations as 
well as with respect to the type and number of potentially
important elementary reaction steps (including decomposition
reactions that determine turnover numbers) may be achieved. 
(ii) Fast electronic structure methods with uncertainty quantification 
warrant high efficiency and reliability in order
to not only deliver results quickly, but also to allow for predictive work.
(iii) A high degree of autonomy reduces the amount of manual human work, processing errors, and human bias.
Although being inherently unbiased, it is still steerable with respect to
specific regions of an emerging network and with respect to the addition of new reactant species.
This allows for a high fidelity of the formalization of some
catalytic process and for surprising in silico discoveries.
In this work, we first review the state of the art in computational catalysis
to embed autonomous explorations into the general field from which it draws its ingredients.
We then elaborate on
the specific conceptual issues that arise in the context of 
autonomous computational procedures, some of which we discuss at an 
example catalytic system.
}

\newpage
\section{Introduction}
\label{sec:introduction}

Catalysis is a key and emergent concept in chemistry: substances are
assigned a special role as they
take part in a reaction but are eventually recovered unchanged after a product
has been formed. It is a chemical insight that such patterns can be discovered in
complex reaction mechanisms. From a quantum chemical point of view, this translates into
producing and then analyzing networks of elementary steps, which map all
(with respect to external conditions such as temperature)
feasible chemical transformations in a sequence of structural changes across a 
Born-Oppenheimer potential energy surface. Understanding catalysis in terms of
such reaction networks can then be a starting point for the design of processes
guided by constraints such as being efficient, cheap, green, and/or sustainable.

Computational catalysis can deliver unprecedented details about catalytic reaction mechanisms~\cite{Norskov2006,Balcells2010,Lin2010,Sautet2010,Norskov2011,VanDerKamp2013,Yang2013a,Thiel2014,Speybroeck2015,Balcells2016,Lam2016,Sperger2016,Vidossich2016,Zhang2016,Romero-Rivera2017,Seh2017,Grajciar2018,Kulkarni2018,Bruix2019,Dubey2019,Vogiatzis2019,Cui2020,Li2020,Funes-Ardoiz2020,Reuter2020,Chen2021a,Chen2021b,Durand2021,Wodrich2021,Hutchings2021,Catlow2021,Lledos2021,Morales-Garcia2021}.
However, a universal theoretical approach toward computational catalysis with generally applicable
algorithms is not available. This can be a handicap for practical applications, especially in view of the growing field of
experimental catalysis with increasingly complex catalyst structures such as
metal-organic frameworks~\cite{Rogge2017,Zhu2017,Bavykina2020,Freund2021},
single-atom catalysts~\cite{Yang2013,Kaiser2020,Samantaray2020},
supported nanoparticles~\cite{Li2020a},
supported organometallic catalysts~\cite{Wegener2012,Coperet2016,Ye2018,Coperet2019a},
binary catalysts~\cite{Chen2014},
encapsulated catalysts~\cite{Worsdorfer2011,Leenders2014,Tetter2017,Jongkind2018},
self-assembling nanostructures~\cite{Azuma2018,Palmiero2020},
nanozymes~\cite{Wu2019},
protein nanocages~\cite{Lv2021},
nucleic-acid catalysts~\cite{Micura2020}, and
artificial (metallo)enzymes~\cite{Davis2019,Arnold2019,Hofmann2020,Chen2020}.

In this work, we first provide a brief overview of the different computational approaches that have been developed
for applications in catalysis and in related fields, before we then
focus on the detailed first-principles modeling of vast elementary reaction networks. It is the very
nature of this complex topic that requires us to touch upon many diverse subjects. While we attempt to provide
a balanced overview with a focus on the most recent developments, we emphasize that a complete literature review
will be impossible to achieve in the context of this work. We therefore consider the numerous references given here as
a starting point for interested readers to dive deeper into the literature of a specific subject. Eventually, we
will focus on automated procedures steered by autonomously working computer (meta)programs. Such approaches
will have a broad future for various reasons to be discussed, but they are still in their infancy. It is for this reason
that we will then consider conceptual aspects of autonomous computational explorations of catalytic systems,
which we then supplement with an explicit example to highlight some of the key issues that need to be mastered.

\section{Computational Catalysis and Mechanism Exploration}
\label{sec:concept}

Considering the complexity of a catalytic process in terms of reaction steps and materials, first-principles modeling
is challenging because of the structural variety and size of the atomistic systems and because of the vast amount
of fine-grained transformation steps that need to be considered. Hence, it is
obvious to exploit existing data first, which has already become a major strategy for the design of new
materials with specific functionality~\cite{Armiento2011,Agrawal2016,Himanen2019,Armiento2020,Yu2021,Blau2021,McDermott2021}.
Vast amounts of data of different origin may be utilized to understand and design
catalytic processes~\cite{Vaucher2021,Schwaller2021}.
A substantial number of publicly accessible databases~\cite{Hachmann2011,Hummelshoj2012,Curtarolo2012,Landis2012,Jain2013,Saal2013,Chung2014,Alvarez-Moreno2015,Kirklin2015,Dima2016,OMara2016,Borysov2017,Draxl2018,Zakutayev2018,Winther2019,Mamun2019,Blokhin2020,Choudhary2020,Talirz2020,Gimadiev2021,Pablo-Garcia2021,Nakata2017,Smith2021,Andersen2021a}
has become available along with software packages encoding general workflows to interact with these databases~\cite{Ong2013,Jain2015,Pizzi2016,Mathew2017,Aagesen2018,Schleder2019,Wheeler2019,Yang2020a,Youn2020,Huber2021,Brlec2021,Wang2021b}.
Screening these data can produce valuable property predictions~\cite{Curtarolo2013,Hachmann2014,Pyzer-Knapp2015,Takahashi2019,Luo2021}.
High-throughput studies can be accelerated by exploiting also surrogate models, \textit{i.e.,}
efficient, empirical models that can produce property predictions such as adsorption energies, albeit less accurately than a first-principles-based model such as density functional theory (DFT)~\cite{Tran2018a}.
Surrogate models can be scaling relationships~\cite{Bligaard2004,Ulissi2017a,Mazeau2021},
physical descriptors~\cite{Xin2012,Zhao2019,Gao2020,Xu2021a}
or machine learning (ML) models trained on physical or structural descriptors~\cite{Ulissi2017,Takahashi2018,Takahashi2018a,Tran2018,Andersen2019,Palizhati2019,Back2019,Deimel2020,Praveen2020,Xu2021,Friederich2021,Li2021,Li2021a,Rosen2021,Andersen2021,Pablo-Garcia2021a,Esterhuizen2021}.
Furthermore, they can be enhanced by stability analysis to save computing time on unstable materials~\cite{Back2020}.
Such fast data-driven hypothesis generation
can then be refined with uncertainty quantification by DFT calculations~\cite{Mortensen2005,Hellman2006,Wellendorff2012,Medford2014,Simm2016,Tran2018a,Tran2020b}.

The application of surrogate models of known uncertainty together with a workflow for high-throughput DFT calculations has been adapted to the evaluation of reaction networks~\cite{Ulissi2017a,FaradayUncertainty2017,Li2018,Simm2018,Stocker2020}.
A small molecular size of reactants, such as the oxidation of CO or the oxidative coupling of methanol, limits the number of possible intermediates during the reaction.
If, in addition, no pronounced structural changes of the catalyst occur during the reaction so that its structure may be regarded as
basically stiff, small chemical reaction networks will emerge that can be considered complete~\cite{Freund2011,Schloegl2015}.
In such a case, a threshold for the maximum molecular size, \textit{e.g.,} number of carbon atoms involved, 
can be chosen to then define a chemical reaction network of all possible elementary steps based on reaction equations~\cite{Ulissi2017a}.

Larger reactants with increased structural degrees of freedom and/or structurally floppy catalysts require
many more elementary steps for reaching a complete reaction mechanism of the catalytic process, typically
much more than can be considered in manual work. Hence, automated procedures are key for the elucidation
of such a complete network in order to uncover all relevant mechanistic steps~\cite{Sameera2016,Vazquez2018,Dewyer2018,Simm2019,Unsleber2020}.
Naturally, definitions of reaction types~\cite{Gu2020} or graph rules~\cite{Margraf2019,Liu2020} have been
exploited for this purpose.
The network of all assumed reaction intermediates on a given surface can then be combined with high-throughput 
quantum chemical calculations and micro-kinetic modeling to compare different existing hypotheses for a reaction mechanism~\cite{Wang2019}.

Constructing a reaction network simply based on viable intermediates on reactants
and considering the catalyst solely as a static partner, onto which these intermediates are adsorbed, is 
mostly limited to flat catalytic surfaces.
Most existing algorithms likely struggle with solid phases that undergo structural rearrangements
during reactions on their surfaces
so that the reaction intermediates significantly differ from their gas phase counterparts;
examples are flexible catalysts such as nanoclusters~\cite{Zhai2017}, anchored organometallic complexes~\cite{Coperet2019}, and reactions that remove and regenerate atoms at a surface~\cite{Mars1954}.
The increased degree of complexity that the direct structural involvement of such catalysts adds to the problem
of the elucidation of catalytic reactions networks for large reactants with a high degree of structural
flexibility highlights an even more pronounced role of automated exploration procedures,
which we, given the diverse nature of potentially catalytic agents, decided to base on 
electronic structure information only~\cite{Bergeler2015,Simm2017,Grimmel2019,Grimmel2021}. This allows us to exploit general heuristic
concepts based on the first principles of quantum mechanics.

Most automated reaction network generation schemes have originally been developed for molecular systems~(see, \textit{e.g.,}~Refs.~\citenum{Maeda2013,Rappoport2014,Kim2014,Wang2014,Bergeler2015,Zimmerman2015,Gao2016,Habershon2016,Guan2018,Kim2018,Grimme2019,Rizzi2019,Jara-Toro2020,Zhao2021}).
The underlying algorithms and concepts range from graph-based rules to the interpretation of the electronic wave function, 
and to \textit{ab initio} molecular dynamics (MD).
However, all these algorithms have the common goal of constructing all possible elementary steps for a given pool of reactants 
by locating the corresponding transition states with first-principles and semi-empirical electronic structure methods.
Whereas they were developed for systems that represent a single phase (typically the gas phase or a solution), 
some of them have also been applied to reactions on metallic surfaces.

The latest release of the graph-based \textit{reaction mechanism generator} (RMG) by Green and co-workers~\cite{Liu2020} features additional graph rules for surfaces, in which the surface is treated as a single graph node with which every other node can form bonds with.
The authors applied this approach to methane dry reforming on Ni~(111)~\cite{Goldsmith2017}, for which their algorithm found many of the reactions of an established mechanism~\cite{Delgado2015}.
However, their approach was limited to predefined reaction types, the adsorption energies were based on literature values or group additivity for missing literature data, and the reaction energy barriers were derived from scaling relationships from the literature.

Zimmerman and co-workers have developed the software \textit{S-ZStruct}~\cite{Jafari2018} for specifically handling surface explorations.
It implements an interface to the \textit{atomistic simulation environment} (ASE)~\cite{Larsen2017} to find adsorption sites and explore reaction paths of adsorbates with their \textit{growing string method} (GSM)~\cite{Zimmerman2015,Jafari2017}.
Maeda \textit{et al.}~have also explored reactions of adsorbates on (111) surfaces~\cite{Maeda2018,Sugiyama2019,Sugiyama2021} 
with their \textit{artificial force induced reaction} (AFIR) approach~\cite{Maeda2021}.
While both approaches, GSM and AFIR, are versatile and general, the application studies 
were limited to low-index surfaces with a completely constrained slab.
Moreover, the adsorption site location of ASE is implemented only for certain surfaces, while more advanced surfaces would require manual definitions~\cite{Jafari2018}.
Owing to the general, atomistic nature of their core algorithm, the AFIR and GSM method, both Maeda \textit{et al.}~\cite{Hatanaka2013,Yoshimura2017,Reyes2020} and Zimmerman \textit{et al.}~\cite{Nett2015,Ludwig2016,Smith2016,Zhao2016,Ludwig2017,Dewyer2017a,Ludwig2018,Rudenko2018,Lipinski2020,Malakar2021,Malakar2021a} have studied homogeneous catalysis more extensively, also incorporating experimental information.
Their algorithms can also be applied in a semiautomatic fashion by steering the exploration into certain branches of the reaction network, either by specifying specific internal coordination transformations or fragments of the molecules that shall be combined or dissociated.
However, this requires extensive knowledge of the software.

In a different approach, Liu and co-workers~\cite{Zhang2017} sampled a reaction on a Cu~(111) surface, namely the water gas shift reaction.
They constrained two of three layers and found the reaction with their enhanced sampling technique called \textit{stochastic surface walking} (SSW).
They applied the SSW technique also for solids~\cite{Guan2015} and more complicated heterogeneous systems~\cite{Ma2019,Ma2019a} by training a neural network on MD data, which is implemented in their \texttt{LASP} software package~\cite{Huang2019}.
Besides surface slabs, also first-principles-based exploration methods have been applied to cluster models of nanoparticles based on graph rules by Habershon \textit{et al.}~\cite{Ismail2019} and with the AFIR approach by Maeda \textit{et al.}~\cite{Song2017,Iwasa2018,Ichino2019}.

A common reference example, that was studied by multiple groups, is the hydroformulation of ethene by the HCo(CO)\textsubscript{3} catalyst with the goal to reproduce the mechanism by Heck and Breslow~\cite{Heck1961}.
This example has been investigated with time-independent calculations by Maeda \textit{et al.}~\cite{Maeda2012}, with an MD based method by Mart\'{\i}nez-N\'{u}\~{n}ez \textit{et al.}~\cite{Varela2017}, and with graph rule based approaches by Habershon \textit{et al.}~\cite{Habershon2016} and by Kim \textit{et al.}~\cite{Kim2018}.

While some proof-of-principle studies on (111) metal surfaces and conformationally limited organometallic catalysts have been conducted, a general software package for autonomous studies of any catalytic system has not been established, yet.
We are developing the software suite \texttt{SCINE}~\cite{Scine2021}, which does not impose heuristics, reaction types, or electronic structure models that are limited to specific chemical systems.
In this article, we introduce the extension of our framework toward general homo- and heterogeneous catalysis, both on a conceptual basis and in terms of first implementations.

We have set out to contribute a general approach to computational catalysis~\cite{Bergeler2015,Simm2017,FaradayUncertainty2017,Unsleber2020}
which is the mapping of chemical reactions on reaction networks in such a way that we can
transcend conventional subcategories of catalysis. To achieve this goal, it is necessary that
all algorithms are agnostic with respect to the type of chemical elements involved and the kind of
chemical process to be considered (in solution, on a surface, in an enzyme, in a metal organic framework or
zeolite, ...). Moreover, the algorithms need to be as stable as possible, requiring operator
interference in an interactive manner~\cite{Haag2014a,Vaucher2016,Heuer2018} only in critical cases where even contiguous
attempts to achieve some target with different algorithmic strategies (such as different 
approaches for transition state searches) have failed~\cite{Haag2014}.

Here, we now focus on automated reaction network constructions for catalysis and
elaborate on the specific challenges which need to be addressed in order to make such constructions feasible for routine application.
For this purpose, the next section first addresses conceptual consideration in the context of catalysis.
Afterwards, we discuss an explicit numerical example to highlight some of the technical challenges as well
as options for their solution.

\section{Conceptual Considerations}
\label{sec:concept}

We first consider conceptual issues that are presented by
problems in catalysis to automated reaction mechanism 
exploration.

\subsection{Identifying catalysis in reaction networks}
\label{sec:concept:identification}
A catalyst is defined as a substance that increases the rate of a chemical reaction.
It is both reactant and product of a reaction and is therefore not consumed~\cite{CatalystIUPAC2021}.
A reaction network that is constructed by automated procedures~\cite{Sameera2016,Vazquez2018,Dewyer2018,Simm2019,Unsleber2020}
and hence not limited with respect to the number and type of reactants (at least in principle)
does not highlight catalytic or autocatalytic cycles that may be embedded within. (Auto)catalysis
is an emergent chemical concept that needs to be discovered in such a network. However, the 
definition of catalysis given above can be turned into an algorithm for its discovery
(see, for example, Ref.~\citenum{Simm2017} for an autocatalytic mechanism detected in a reaction
network of the formose condensation reaction).

\begin{figure}[H]
	\begin{center}
	   \includegraphics[width=0.6\linewidth]{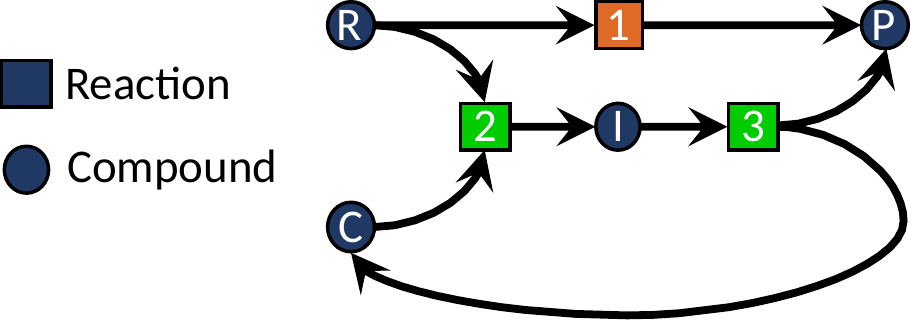}
   \end{center}
   \caption{\label{fig:catalyst_definition}\small A minimal reaction network including a reactant $R$, catalyst $C$, intermediate $I$, and product $P$. The orange colored reaction features a larger barrier height than the green colored reactions.}
\end{figure}

Since a vast reaction network of elementary reaction steps
is \textit{a priori} agnostic with respect to our understanding of
some of its substructures as being catalytic, their identification 
follows \textit{a posteriori} by searching for properties given in the definition of 
a catalyst: 1) An individual molecule or atomistic ensemble (such as a surface) is identified
to take part in a reaction, but is recovered at another position in the network. 
2) The other reactants and products of this reaction are found to be connected by a set of different elementary
steps somewhere else in the network. 3) Then, one may be able to extract two net reaction rates for
both reactions (one with the entity that emerges unchanged from the reaction and one without such an entity). 
4) If the reaction rate with the eventually recovered species is significantly faster than the one
without, this species will most likely be a catalyst --- obviously, the increase in rate must be
significant for a catalyst in order to distinguish its role from that of a pure spectator molecule
such as a solvent molecule.
A minimal reaction network is depicted in Fig.~\ref{fig:catalyst_definition}, where the compound $R$ can either react uncatalyzed to $P$ in reaction 1 or via the reactions 2 and 3 enabled by the catalyst $C$.
A discovery of (auto)catalytic processes in this way is relevant mostly for exploratory
studies of vast reaction networks, for which hardly any or no information
is available at the start of the exploration. In practice, the problem is often simplified by
the fact that one may know the (standard or a class of) catalyst structures to be investigated (and also of the
chemical reaction that is to be catalyzed). A catalytic cycle can then be explored in a
straightforward manner and directly compared with the reaction that lacks the catalyst as a reactant 
(typically in two different explorations conducted in parallel). This procedure is clearly more
target-oriented and allows for catalyst design (by modification and subsequent refinement of
structures in a catalytic cycle; see below) as well as for the evaluation of the catalytic potential
by direct energy-based comparison with the catalyst-free reference network.

\subsection{Calculation of well-established diagnostics from reaction networks}
\label{sec:concept:diagnostics}

Given a vast reaction network that includes an identified catalyst, the question remains, how the catalytic mechanism can be understood and quantified from this network.

Micro-kinetic modeling of the network, \textit{e.g.,} by solving a Markovian master equation based on state and transition
probabilities~\cite{Froment2005,Glowacki2012,Sabbe2012,Stamatakis2012,Stamatakis2014,Gusmao2015,Oliveira2016,Reuter2016,Park2019,Motagamwala2021},
preferably accounting for first-principles-derived uncertainties in these probabilities~\cite{Sutton2016,FaradayUncertainty2017,Proppe2019}, is desirable.
However, this is computationally demanding for vast reaction networks, especially if several reaction networks should be compared with one another.
Therefore, some limitations of the network or approximations for the kinetic analysis are commonly introduced (see, for instance, Refs.~\citenum{Campbell2017,Maffei2021}).

Instead of constructing reaction networks based on heuristic rules and then conducting a kinetic analysis on the whole network, one may 
explore the reaction network based on quantum chemical methods
with a continuously running  kinetic analysis on the fly as a guide. Such a kinetics-driven steering of the exploration process can exploit the calculated barrier heights obtained so far to determine those nodes that accumulate concentration and are therefore the key nodes for further network exploration in the next step~\cite{Simm2017,Proppe2019}.

Two general experimental metrics for the effectiveness of catalysts are the \textit{turnover number} (TON) and the \textit{turnover frequency} (TOF).
However, their definitions may vary for different types of catalysis such as biocatalysis, homogeneous catalysis, and heterogeneous catalysis~\cite{Bligaard2016}.
We take the TON to be a quantitative measure for the stability of a catalyst against deactivation reactions and the TOF as a measure for the efficiency of a catalyst.

We first define the TOF as the number of catalytic reaction cycles $N^c$ accomplished per time $t$
\begin{equation}
    \text{TOF} = \frac{N^c}{t},
\end{equation}
which may be obtained numerically by micro-kinetic modeling of a reaction network or analytically by identifying the catalytic cycle within a network and applying the \textit{energetic span} model \cite{Kozuch2006,Kozuch2008,Kozuch2010,Kozuch2011}.

Experimentally, this quantity must be normalized by some measure for the amount of catalyst available. 
One may compare experimental results and theoretical predictions based on first-principles networks based on relative theoretical TOFs rather than on absolute TOFs for reasons discussed later.

In heterogeneous catalysis, the TOF is commonly replaced with the \textit{site time yield} \cite{Boudart1995}, which is normalized with the number of active sites on the catalyst that may be approximated by the number of adsorbing gas molecules in a separate experiment.
We do not need to consider such a normalization, because a complete chemical reaction network at full atomistic resolution would include a catalytic cycle for each and every individual active site. Hence, one would obtain a theoretical TOF per site and then may average over all sites afterwards, if desired.

Theoretical TOFs are often determined in the framework of transition state theory (TST)~\cite{Eyring1935}, which connects the reaction rate $k_i$ with the activation free energy $\Delta G^{\ddagger}_i$
\begin{equation}
k_i = \frac{k_B T}{h} e^{-\beta \Delta G_i^{\ddagger}}
\end{equation}
with Plank's constant $h$, temperature $T$ and $\beta$ defined as the inverse product of the Boltzmann constant $k_B$ 
and $T$, \textit{i.e.,} $(k_B T)^{-1}$.
In the framework of TST, Kozuch and Shaik derived the \textit{energetic span} model~\cite{Kozuch2006,Kozuch2008,Kozuch2010,Kozuch2011}, which allows one to calculate the TOF from the absolute Gibbs energies of all transition states $G^T_i$ and intermediates $G^I_j$ and the relative Gibbs energy $\Delta G_r$ of the catalytic cycle of $N$ steps
\begin{equation}\label{eq:tof}
	\text{TOF} = \frac{k_B T}{h} \frac{e^{-\beta \Delta G_r} - 1}{\sum_{i,j = 1}^{N} e^{\beta (G^T_i - G^I_j - \delta G_{i,j})}}, \qquad \delta G_{i,j}
	\begin{cases}
	\Delta G_r,& \text{if } i > j\\
	0,& \text{if } i \leq j.
	\end{cases}
\end{equation}
This general expression can be approximated in terms of two crucial concepts, the \textit{TOF determining transition state} (TDTS) and \textit{TOF determining intermediate} (TDI)~\cite{Kozuch2011}:
\begin{equation}
\text{TOF} \approx \frac{k_B T}{h} e^{-\beta \delta E}, \quad \delta E
\begin{cases}
T_{\text{TDTS}} - I_{\text{TDI}},& \text{if TDTS appears after TDI}\\
T_{\text{TDTS}} - I_{\text{TDI}} + \Delta G_r,& \text{if TDTS appears before TDI}.
\end{cases}
\end{equation}
The two states, TDTS and TDI, maximize the \textit{energetic span} $\delta E$ of a catalytic cycle.
The reliability of this approximation depends on the \textit{degree of TOF control}~\cite{Kozuch2011} of TDTS and TDI.

By virtue of the \textit{energetic span} model, the activity of a catalytic reaction cycle within a chemical reaction network can be directly estimated~\cite{Kozuch2015}. Additionally, crucial states and steps within a reaction mechanism can be identified.
Kozuch and Shaik showed that comparisons of calculated TOFs are quantitatively reliable due to error cancellation, while absolute rate estimates are difficult to predict
due to the exponential amplification of an error in the Gibbs energy~\cite{Kozuch2011}.

The robustness of relative rate comparisons allows also for reliable estimates of the proportion of occurring catalytic reactions and degradation reactions, which allows to calculate the TON.
For this, we define catalytic reactions $r_i^c$, \textit{i.e.,} a single reaction or series of reactions, for which a species has been
identified to act as a catalyst and is therefore recovered after the reaction.
We also define degradation reactions $r_j^d$, for which the catalyst is solely a reactant and not recovered, and we define degradation reactions $r_k^d$, which also consume the catalyst, but require an intermediate of a catalytic reaction $r_i^c$ as reactant.
Note that 'reaction' here refers to a sequence of elementary steps. In other words,
if a catalytic reaction consists of multiple elementary steps, which is typically depicted as a cycle, it is solely one $r_i^c$ in our definition.

In view of the data that are available for a reaction network of elementary steps, it would be convenient to define a 'turnover efficiency'
as a measure for the TON that can be obtained as the ratio 
of the total probability for product molecular production and the total probability for catalyst decomposition.
Naturally, such probabilities are given by the net rate constants for sequences of elementary steps that 
either lead to product molecules or to catalyst decomposition.
Accordingly, we may introduce such a TON as the ratio of the sum of rate constants $k_i^c$ of all catalytic reactions and the sum of rate constants $k_j^d$ of all degradation reactions
\begin{equation}\label{eq:ton}
\text{TON} = \frac{\sum_i k_i^c (r_k^d)}{\sum_j k_j^d}.
\end{equation}
As indicated in the numerator, the rate constant of the catalytic reaction(s) $k_i^c$ is, among other quantities, a function of the degradation reactions $r_k^d$ that branch off the catalytic cycle -- which only affect catalytic reaction $r^c_i$ -- while the degradation reactions $r_j^d$ disconnected from any catalytic cycle affect all $r^c_i$ and lower the total TON.
Generally, $k_i^c$ can be approximated by the TOF, but due to this additional consideration of $N_d$ degradation reactions with $\Delta G_k^{\ddagger}$ Gibbs energy barriers, eq.~(\ref{eq:tof}) must be slightly altered to read
\begin{equation}\label{eq:tof-mod}
k_i^c = \frac{k_B T}{h} \frac{e^{-\beta \Delta G^c_i} - 1}{\sum_{a,b = 1}^{N} e^{\beta (G^T_a - G^I_b - \delta G_{a,b})} \sum_{k = 1}^{N_d} e^{-\beta \Delta G_k^{\ddagger}}}, \qquad \delta G_{a,b}
\begin{cases}
\Delta G_r,& \text{if } a > b\\
0,& \text{if } a \leq b.
\end{cases}
\end{equation}
The TON can then solely be expressed in terms of energies as
\begin{equation}\label{eq:ton_energies}
\text{TON} = \frac{\sum_i 
	\frac{e^{-\beta \Delta G^c_i} - 1}{\sum_{a,b = 1}^{N} e^{\beta (G^T_a - G^I_b - \delta G_{a,b})} \sum_{k = 1}^{N_d} e^{-\beta \Delta G_k^{\ddagger}}}, \qquad \delta G_{a,b}
	\begin{cases}
	\Delta G_r,& \text{if } a > b\\
	0,& \text{if } a \leq b.
	\end{cases}}
{\sum_j e^{-\beta \Delta G_j^{\ddagger}}}
\end{equation}
This allows us to calculate the stability of a catalyst against decomposition. However, this is hardly done in experimental research~\cite{Jones2010} and neither in computational research due to the complexity of finding all relevant degradation reactions. A mitigation of this
problem is, in fact, the autonomous exploration of elementary steps based
on automated first-principles procedures, which can deliver huge networks
of complex reactions that may be considered complete after a certain exploration depth has been reached.

\subsection{Autocatalysis}
\label{sec:concept:autocatalysis}
The simplest definition of autocatalysis is given by a (series of) elementary step(s), in which a product $X$ catalyzes its own creation~\cite{Schuster2019}.
\begin{equation}\label{eq:autoreaction}
\ce{A + X -> 2X}
\end{equation}
Due to the \textit{nonlinear chemical dynamics}~\cite{Sagues2003} (such as oscillations) that autocatalysis can cause, it has attracted 
little interest by the chemical industry until recently~\cite{Schuster2019}.
Accordingly, the topic has received much attention in origin of life studies~\cite{Blackmond2009,Weissbuch2011,Meyer2012,Vaidya2012}, since autocatalysis can be connected to replication, which is essential for the development of complex living organisms. It might also be the cause of homochirality of all amino acids within all living beings on Earth~\cite{Hein2012}.
Recently, autocatalytic self replication has been developed and studied in synthetic chemical systems~\cite{Mondloch2012,Virgo2016,Semenov2016,Kosikova2017}.

On a theoretical basis, autocatalytic reaction networks have been studied as a basis of the origin of life by Eigen~\cite{Eigen1971}, Kauffman~\cite{Kauffman1986}, and Steel~\textit{et al.}~\cite{Steel2000,Hordijk2004}.
Steel~\textit{et al.}
developed the \textit{reflexively autocatalytic food generated} (RAF) network model, that was also applied in the study of metabolic pathways~\cite{Sousa2015} based on slightly modified or grouped reactions stored in the UniProt database~\cite{UniProt2019} to fit the RAF model.
Note, however, that Andersen~\textit{et al.}~criticized the RAF model for assuming that every reaction within a chemical reaction network is catalyzed, which is unlikely~\cite{Andersen2021b}.
Instead, Andersen~\textit{et al.}~developed a rigorous definition of autocatalysis in chemical reaction networks by describing the network as a directed hypergraph and the autocatalytic reaction as an integer hyperflow~\cite{Andersen2019a} based on reactions derived from graph rules.
However, they noted that a sole definition by hyperflows is most likely not sufficient and will need complementary constraints in order to detect autocatalytic cycles in arbitrary chemical reaction networks~\cite{Andersen2021b}.

Such algorithms, which avoid computationally expensive numerical kinetic simulations, are required and cannot be circumvented with a straightforward identification strategy solely based on thermodynamic criteria as outlined in section~\ref{sec:concept:identification}.
For example, the corresponding uncatalyzed reaction of eq.\ (\ref{eq:autoreaction}), which can be formulated as
\begin{equation}
\ce{A -> X}
\end{equation}
might simply not exist or impose such high barriers that it cannot be located with standard algorithms.
Without such points of references, which are missing in experimental data of biological systems, for which most definitions and algorithms 
discussed in this section had been developed, autocatalysts can only be identified and distinguished from bystander molecules based on kinetic analyses.

\begin{figure}[H]
	\begin{center}
	   \includegraphics[width=0.6\linewidth]{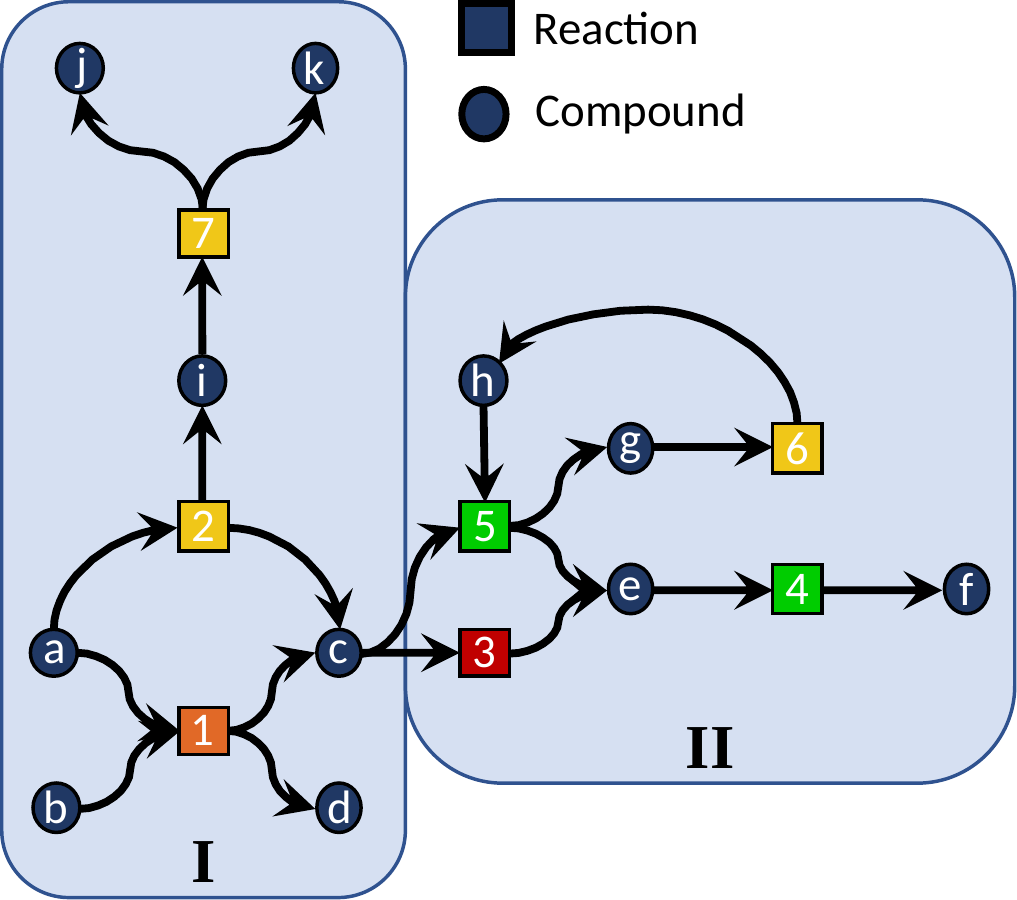}
   \end{center}
   \caption{\label{fig:autocatalysis}\small A reaction network including the autocatalytic reaction 5. Light-green reactions have the lowest reaction barrier heights, followed by yellow, orange, and dark-red (indicating the largest barriers).}
\end{figure}

Many theoretical models also construct chemical reaction networks solely with graph rules and do not take into account different reaction barriers and conformers.
If the exploration of a network is based on first-principles calculations in such a way that all elementary steps are mapped out, the detection of autocatalysis requires micro-kinetic modeling of the reaction network.
However, if one restricts the exploration by constraints that do not allow for the
passing of barriers of a given height (or similarly by explicit kinetic modeling), the detection of autocatalytic paths becomes much more difficult, especially for compounds which can only be formed by an autocatalytic reaction.
The issue is that a product, which might act autocatalytically and, therefore, decreases the barrier(s) of the reaction(s) necessary for its own creation, might never be found.
A minimal example is depicted in Fig.~\ref{fig:autocatalysis}, where compound $h$ acts autocatalytically.
In a first-principles-based exploration of this network starting from $a$ and $b$, the network would never discover the region \MakeUppercase{\romannumeral 2} leading to the favored product $f$, but would, instead, stay in region \MakeUppercase{\romannumeral 1} and 
wrongly predict the compounds $j$ and $k$ as the major products.

The crucial question then is how one can account for this issue in the automated exploration of a chemical reaction network? For known autocatalytic motifs, a viable option would be the systematic trial exploration of such a motif. An example is acid catalysis in the context of ester hydrolysis (see, for instance, Ref.~\citenum{Bissette2013} for a detailed description and further examples). If many exhaustive catalytic reaction networks become available in the future so that sufficient amounts of data are available, one may extract patterns for the onset of autocatalytic pathways with machine learning models.
Unfortunately, all of this would include a heuristic bias on known chemical phenomena and further research is required to identify 
truly exploratory first-principles-based approaches.

\subsection{Catalyst Design}
\label{sec:concept:design}

Many optimization and design strategies for more stable or active catalysts have been developed for specific fields such as
biocatalysis~\cite{Arnold2001,Jiang2008,Siegel2010,Hilvert2013,Kiss2013,Zastrow2013,MunozRobles2015,Zhang2015,Alonso-Cotchico2020,Bunzel2021},
homogeneous catalysis~\cite{Maldonado2010,Robbins2011,Raugei2015,Doney2016,Wheeler2016,Poree2017,Lu2019,Foscato2020,Rinehart2021,DosPassosGomes2021,Nandy2021},
or heterogeneous catalysis~\cite{Norskov2009,Greeley2016,Personick2016,Jimenez-Izal2018,Zhao2020,Wang2021,Guo2021}.
In these strategies, the activity of a catalyst is judged on various physical descriptors. For our discussion here, it is important to recall that
a chemical reaction network of elementary steps is a universal means for studying a catalytic reaction: it encodes all information for understanding the catalytic process \textit{in toto} (including deactivation processes and side reactions). Once the reaction states that are key for a catalytic process (\textit{e.g.}, those that determine TON and TOF) have been identified, they can become a target for catalyst optimization and even for \textit{de novo} design. Note that the uncatalyzed reaction itself is already a viable starting point as its network contains those steps that require a catalyst to decrease high reaction barriers. As such, the network provides atomistic structural information about where and possibly also about how to introduce structural changes and potentially catalytic reagents. Naturally, any structural change introduced at some node of the network will then require a re-evaluation of the whole network in order to probe the viability of previously found elementary steps, to find new ones, and to assess the resulting activation (free) energies. While this is a computer time demanding task, tailored optimization strategies that target specific structure-property relationships may decrease the computational burden.

In general, it will neither be feasible nor sensible to automatically explore a complete reaction network from scratch for a large number of potential catalyst candidates, prohibiting high-throughput screening for catalysts based on networks of elementary steps.
Instead, the comparison between different catalysts should happen on the basis of network inheritance in order to be efficient. 

First, the chemical reaction network may be explored with one specific catalyst, \textit{e.g.,} the known reference catalyst that should be improved.
To increase the efficiency of the exploration, this catalyst should be generic in the sense that its structure should not possess 
unnecessarily costly elements; \textit{i.e.}, those that can be expected to be spectator residues for the catalytic process itself,
but would increase the computational time significantly.
A typical example are substituents with large conformational freedom that can be expected to play hardly any role in the catalytic process itself, but are required for different purposes (\textit{e.g.,} solubility or preventing catalyst dimerization). Such structural elements may be discarded for the generic catalyst for which it is then much easier to generate a complete reaction network as it will not suffer from a combinatorial explosion of conformers. 
However, crucial misrepresentations of the catalyst, which foundamentally change the reaction mechanism, have to be avoided~\cite{Harvey2019}. 

In a subsequent step, one may re-introduce substituents (also for the purpose of catalyst design) in a step-
or shell-wise fashion, possibly aided by ML approaches~\cite{Cordova2020,Chen2021d}. The generic reaction network can then serve as an efficient starting point, allowing for a fast re-evaluation of its nodes with the new catalyst structure and a search for new elementary steps. 

\begin{figure}[H]
	\begin{center}
	   \includegraphics[width=\linewidth]{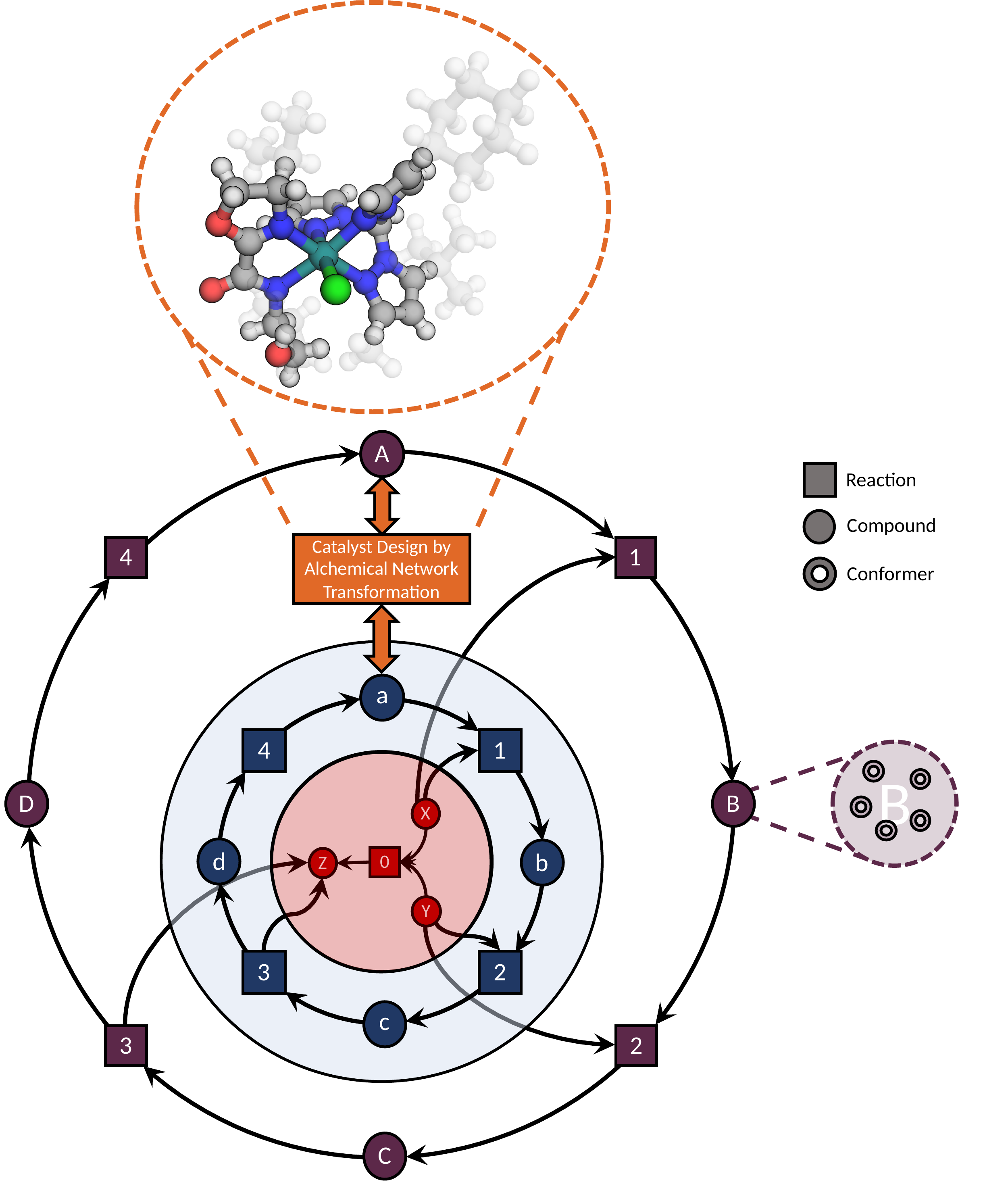}
   \end{center}
   \caption{\label{fig:catalyst_design}\small A schematic reaction network depicting the uncatalyzed reaction 0 of $X$ and $Y$ to $Z$ (red). The same chemical reaction can also be found catalyzed by a minimal catalyst $a$ in the series of reactions 1-4 (in blue). This minimal cycle can then be exploited for catalyst design by systematically exchanging ligands (or substituents or central metal ions) of the catalyst, which is schematically depicted in the circle at the top. The modified reaction barriers for 1-4 based on the new catalyst $A$ (in purple) can then be explored within the reaction network.}
\end{figure}

Alternatively, the main catalytic entity -- in most cases a metal or a certain structural motif -- can also be substituted on a network level to study different candidates. The simplest case is that of a 'transmutation' where the metal in all structures of
generic network is simply exchanged by another one, for
which a homologous metal or an isoelectronic metal fragment are suitable candidates (consider, for example,
replacing Ru by Fe or Co$^+$) as depicted in Fig.~\ref{fig:catalyst_design}.

In this way, information about the catalytic process is inherited in such a way that computational costs are efficiently reduced and the emerging ancestry can enhance the conceptual understanding of the catalytic system.

Since this is a direct approach, in which a molecular structure is given and its property is calculated, high-throughput virtual screening (HTVS) must be conducted to search for a better catalyst in a systematic way.
However, even with an efficient HTVS approach, it is hardly possible to visit a sufficiently large fraction of the chemical space due to its sheer size~\cite{Kirkpatrick2004,Reymond2015}.
Therefore, a wise selection of compounds and materials of this space has to be made depending on the design target.

The key problem is that quantum and classical mechanics allow us to predict a molecular property or function for a molecular structure given. The inverse direction, \textit{i.e.,} from a desired function to a molecular structure that exhibits this function, is mathematically ill-defined for various reason (\textit{e.g.,} in quantum mechanics all dynamical degrees of freedom (such as coordinates) are integrated out when expectation values or response properties are calculated). However, one may hope to develop inverse approaches for specific goals as certain properties of these goals may be exploited to alleviate the problem.

Accordingly, \textit{inverse design} strategies attempt to predefine a specific target property and then construct the corresponding ensemble of structures that feature this property.
Many approaches for such algorithms exist and have been discussed in general reviews~\cite{Weymuth2014a,Zunger2018,Freeze2019}
and reviews focusing on ML approaches for \textit{inverse design}~\cite{Sanchez-Lengeling2018,VonLilienfeld2020,Lu2021,Pollice2021}.

For example, we have proposed the inverse-design approach \textit{Gradient-driven Molecule Construction} (GdMC)~\cite{Weymuth2013,Weymuth2014,Krausbeck2017}, which targets design of new catalysts by sequentially constructing metal fragments that stabilize structurally activated small molecules in intermediates through reduced structure gradients on all atoms.
In another approach, Hartke and co-workers have combined optimizations of minimum energy reaction paths in an electric field of point charges with global optimization techniques in their \textit{Globally Optimized Catalyst} 
scheme~\cite{Dittner2018,Dittner2020} and have further improved on it in a quantum-mechanical molecular-mechanical composite
approach~\cite{Behrens2021}.

ML had a considerable impact on the field of \textit{inverse design} in recent years as it allows for learning
structure-property relationships, which can then be employed to generate structures based on a given property.
Especially deep generative models have been demonstrated to be successful across multiple chemical problems
ranging from drug discovery~\cite{Gomez-Bombarelli2018,Boitreaud2020,Lim2020} to materials design~\cite{Yao2021,Pathak2020,Kim2020}.
A combination of such models with genetic algorithms is also possible~\cite{Nigam2021b}.
For this endeavor to be successful, it was necessary to improve on the representation of chemical structures~\cite{Krenn2020,Nigam2021}
and desired properties~\cite{Meyer2018}.
It was also shown that the new concept of alchemical chirality~\cite{Rudorff2021} might allow one
to draw direct energy relations across the chemical compound space to accelerate design processes.

Hence, many strategies have been developed for the design of molecules with specific properties. It can be 
expected that catalyst and process design by computational catalysis will continue to strive for novel as well as routinely applicable
design protocols.  

\section{Computational Considerations}
\label{sec:results}
Because of the numerous elementary steps involved in catalytic processes and the fact that changes in structural composition point to new networks of elementary steps, the computational burden is truly intimidating and smart procedures are required to keep it feasible in principle, but also in view of the environmental footprint of high-performance computing campaigns.
In this section, we therefore turn to a discussion of the computational resources required for autonomous first-principles-based explorations of homogeneous and heterogeneous catalysis that allow for an understanding on the basis of reaction networks. Clearly, the computational resources required will depend on the methodology chosen. Here, we rely on our methodology 
in order to give an idea of the magnitude of computational effort that is to be invested 
in autonomous first-principles-based explorations. Our computational methodology is detailed in the appendix.

\subsection{Resource estimates for automated explorations of homogeneous systems}
\label{sec:ressource_estimates_homo}

A chemical reaction network can be constructed solely based on initial reactants as input. Starting from these structures, all elementary steps can be identified -- at least in principle -- by letting algorithms search for new local minima starting from the given ones on the respective Born-Oppenheimer hypersurfaces. Newly found minima, which correspond to long- or short-lived intermediates in a reaction network, become new starting points for further exploration in this rolling approach.

A key part of autonomous explorations are automated procedures that allow for the identification of elementary reaction steps with associated transition states.
For instance, with our \texttt{Chemoton} exploration software, possible elementary steps are probed based on reaction coordinates defined for active sites identified within molecules. In principle, every atom (or group of atoms) in a molecule may function as an active site, an assumption that allows one to map out a reaction network that is as complete as possible. However, this will often not be feasible and
so protocols are put in place that reduce the number of potentially
relevant sites to those that might be active under reaction conditions. Our strategy so far has been to base this selection 
process on rules that may be derived for any molecular system and that are therefore not bound to specific compound classes. Accordingly, we introduced first-principles heuristics as a way to extract conceptual information on reactivity from the electronic wave function~\cite{Bergeler2015,Simm2017,Grimmel2019,Grimmel2021}. Note that it is not required to make a precise prediction on what atoms may react in some intermediate. Instead, it will already be sufficient to identify with certainty those sites that will not react for diminishing the computational burden.

In a brute-force approach, one possible ansatz is to define an inter- or intramolecular reaction coordinate as a push (or pull) of
reactive centers, which in turn can be defined as the geometric center of one or more reactive sites. This then allows one to
enumerate all possible inter- and intramolecular reactions.
\texttt{Chemoton} probes potential reaction coordinates with so called \textit{Newton trajectories}, for details see the appendix.
An exploratory reaction coordinate can be defined as the vector between two geometric centers of lists of active sites.
A geometric center is defined by a number $a$ of active sites, with $a \ge 1 \wedge a \le n_i$ and $n_i$ being the number of nuclei in a reactant.
The second center is then defined by a different list of $b$ active sites.
For intramolecular reactions the reaction coordinate is simply the vector between the centers, while intermolecular reactions require an additional vector for each combination of active sites and angle between these vectors to construct such an exploratory reaction coordinate.
For each combination of $a$ active sites there exists an infinite number of $d_a$ possible vectors and $\rho_a$ possible rotamers, which are reduced in \texttt{Chemoton} by discretization of the rotational angle to a finite number based on steric criteria and a fixed number of rotamers.
To estimate the scaling of such a brute-force approach, we limit possible intermolecular elementary steps to bimolecular reactions.
In a reaction network of $m$ compounds found at a given point in time, a number of $n_{ci}$ structures per compound $i$ with $n_i$ nuclei each
allows us to estimate the number of possible reaction trials $r$ as
\begin{equation}
r = \underbrace{\sum_{i=1}^{m} \sum_{j=i}^{m} n_{ci} n_{cj} \sum_{a=1}^{n_i} \sum_{b=1}^{n_j} d_a d_b \rho_a \rho_b {n_i\choose a}{n_j\choose b}}_{\text{intermolecular reactions}}
+ \underbrace{2\sum_{i=1}^{m} n_{ci} \sum_{a=1}^{\left \lfloor{\frac{n_i}{2}}\right \rfloor} \sum_{b=a}^{n_i-a} {n_i\choose a+b}}_{\text{intramolecular reactions}}.
\end{equation}
The factor 2 for intramolecular reactions stems from the possibility of either associative or dissociative reactions, while intermolecular reactions can only be associative, albeit they can still generate multiple products.
We emphasize that the above equation solely rests on combinatorial considerations that ignores all chemistry knowledge.
It is obvious that activating chemical knowledge will dramatically decrease the number of options -- the question is
how this can be achieved in a way that is so general that it works for any sort of atomistic system, ranging from
molecules to molecular aggregates and eventually to surfaces and composite materials.

Note that $r$ represents only the number of elementary step trials (\textit{i.e.}, attempts to identify an elementary step) and not the number of successful elementary steps, because chemical reactions will not be possible for every combination of nuclei.
Nevertheless, $r$ grows factorially with $n_i$ 
and quadratically with $m$, because any intermediate or reactant can react with any other one of the network. This quickly
becomes unfeasible for a large system, which is why pruning (for instance, through first-principles heuristics) will be necessary for the elementary step trials even in exhaustive reference reaction network explorations.

For our resource estimates, we introduce the assumption of maximally combining pairs of active sites ($a\le  2 \wedge b\le2$) 
for intermolecular reactions, and only pairs of single active sites
($a=b=1$) for intramolecular reactions, which then leads to
\begin{equation}\label{eq:scaling}
r = \underbrace{\sum_{i=1}^{m} \sum_{j=i}^{m} n_{ci} n_{cj} \sum_{a=1}^{2} \sum_{b=1}^{2} d_a d_b \rho_a \rho_b {n_i\choose a}{n_j\choose b}}_{\text{intermolecular reactions}}
+ \underbrace{\sum_{i=1}^{m} n_{ci} (n_i^2 - n_i)}_{\text{intramolecular reactions}}.
\end{equation}
This reduces the scaling behavior to $\mathcal{O}(m^2n_i^4)$. If we assume $m \gg n$ -- i.e., there are far more stable intermediates
in the network than, on average, atoms in each of the intermediates, which is the case for most molecular networks --, 
then the scaling will become quadratic.

Next, we impose restrictions based on graph distances $\delta_{AB}$, which can be determined from Mayer bond orders~\cite{Mayer1983} and our \texttt{Molassembler} library~\cite{Molassembler2020,Sobez2020}, which is part of the
\texttt{SCINE} project.
The graph distance $\delta_{AB}$ is defined as the number of bonds that one passes when proceeding from nucleus $A$ to nucleus $B$ in the 
molecular graph.
Elementary step explorations $r_{\{A,B\}-\{C,D\}}$ are defined with a reaction coordinate constructed between the active sites $A$ and $B$ and the active sites $C$ and $D$.
We limited the number of $r_{\{A,B\}-\{C,D\}}$ depending on the explored reaction type
\begin{align}
\text{intermolecular association:  } &r_{\{A,B\}-\{C,D\}} &&\Leftrightarrow \delta_{AB}=1 \wedge \delta_{CD}=1\\
\text{                             } &r_{\{A\}-\{C,D\}} &&\Leftrightarrow \delta_{CD}=1\\
\text{                             } &r_{\{A,B\}-\{C\}} &&\Leftrightarrow \delta_{AB}=1\\
\text{intramolecular association:  } &r_{\{A\}-\{B\}} &&\Leftrightarrow \delta_{AB}=5 \vee \delta_{AB}=6 \\
\text{intramolecular dissociation: } &r_{\{A\}-\{B\}} &&\Leftrightarrow \delta_{AB}=1.
\end{align}
Additionally, we applied a symmetry analysis to reduce the number of unique active sites and only considered further explorations 
for compounds, which were accessible by reactions with barrier heights below 200~kJ~mol$^{-1}$.
Moreover, to properly sample the remaining elementary steps, we considered two rotamers per reactant ($\rho_a = 2$) and multiple directions of attack ($d_a \ge 1$), where multiple local minima in steric hindrance around the active site were present.

\begin{figure}[H]
	\begin{center}
		\includegraphics[width=0.8\linewidth]{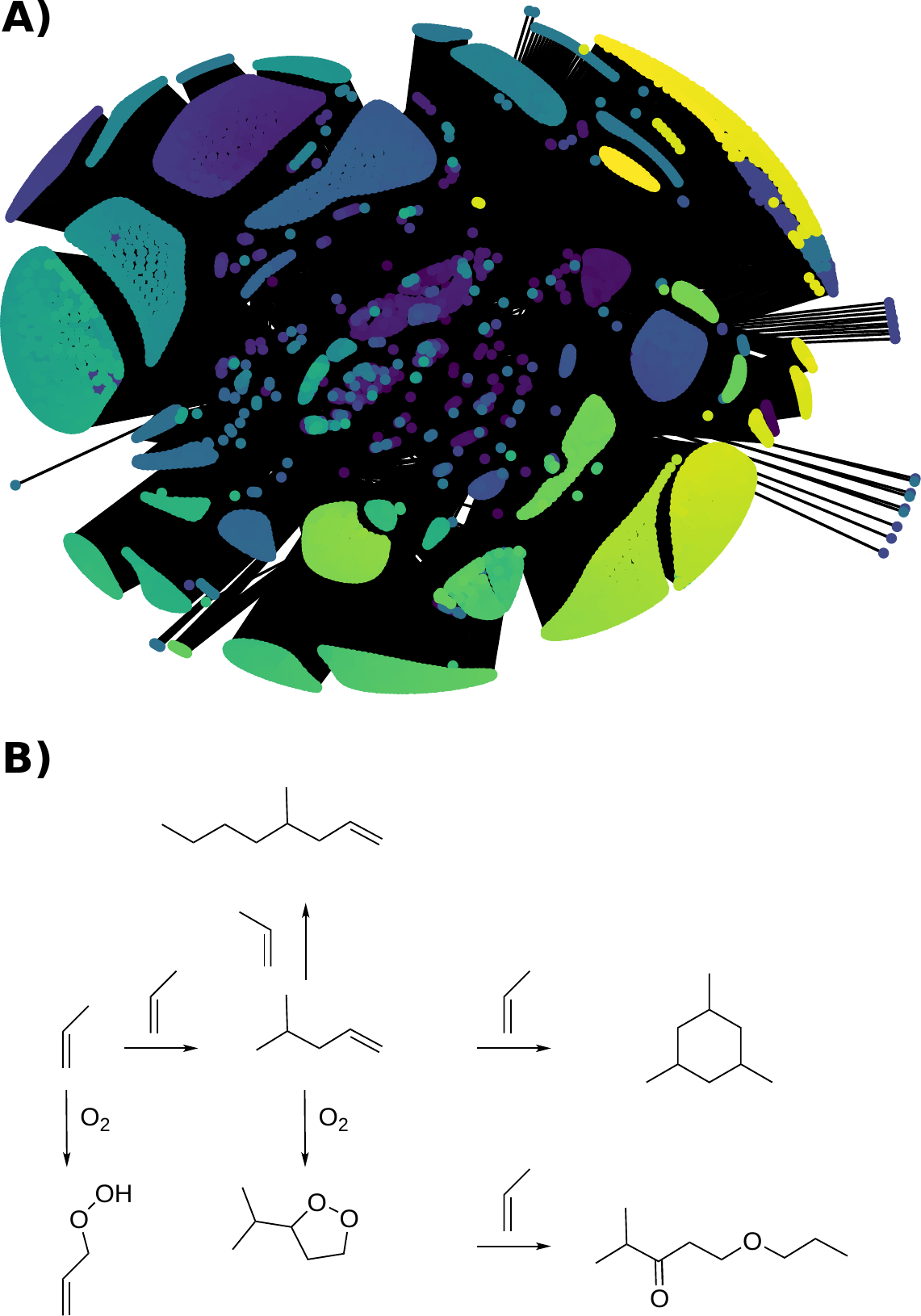}
	\end{center}
	\caption{\label{fig:network}\small \textbf{A)} All compounds in our reaction network connected with lines corresponding to reactions. The compounds are colored according to their order of discovery from violet to yellow. \textbf{B)} Examples of some of the first reactions in the network.}
\end{figure}

First, we constructed from first principles a broad reference reaction network without a catalyst.
Such a network allows us to estimate the scaling effects of the restrictions imposed on the explored elementary steps.
As an example, we selected propylene and molecular oxygen, which already allowed us to construct a broad reaction network from first principles as shown in Fig.~\ref{fig:network}.
This illustrates the potential scope of reaction networks for small systems.

For the uncatalyzed reference, we explored the reaction network starting from propylene and molecular oxygen with
GFN2-xTB~\cite{Bannwarth2019,Bannwarth2021}. We stopped the exploration after $\approx 3 \cdot 10^6$ elementary step trials carried out in a total computing time of 5,775 CPU days and $\approx 1.4 \cdot 10^7$ elementary step trials still remaining.
This resulted in 4,218 compounds, 909 of which are accessible with reaction barrier heights below 200~kJ~mol$^{-1}$. The 4,218 compounds include a total of 1,185,893 individual optimized minimum energy structures that
are connected by 587,752 transition state structures in elementary steps, which were grouped into 6,323 reactions.
For the exploration we set an upper limit in terms of element composition of C$_{10}$H$_{22}$O$_{7}$ and the heaviest compound in our explored reaction network is C$_{9}$H$_{18}$O$_{4}$.
The exploration required a total of $\approx 2.9 \cdot 10^9$ 
single-point calculations, which, for the sake of comparison,
corresponds to a total runtime of $\approx 1.45$~$\upmu$s of a continuous MD simulation with a timestep of $0.5$~fs. 

The most straightforward solution to reduce the number of elementary step trials is a pre-selection based on reactivity descriptors (\textit{e.g.}, first-principles heuristics; see above), which was deliberately {\it not} considered in our reference network.
The fact that we did not activate such a selection/exclusion schemes for the assignment of active sites to be subjected to
elementary search trials can also be observed in the low success rate $\sigma$ of only 22~\% in our brute-force approach.
Furthermore, we could have restricted the number $m$ of intermediates to be considered as reactants by exploiting
some measure for their lifetime. For instance, an intermediate connected to other low-energy intermediates by low barriers
will be short-lived and may be excluded from the set of $m$ reactants to be considered.

\begin{figure}[H]
	\begin{center}
	   \includegraphics[width=\linewidth]{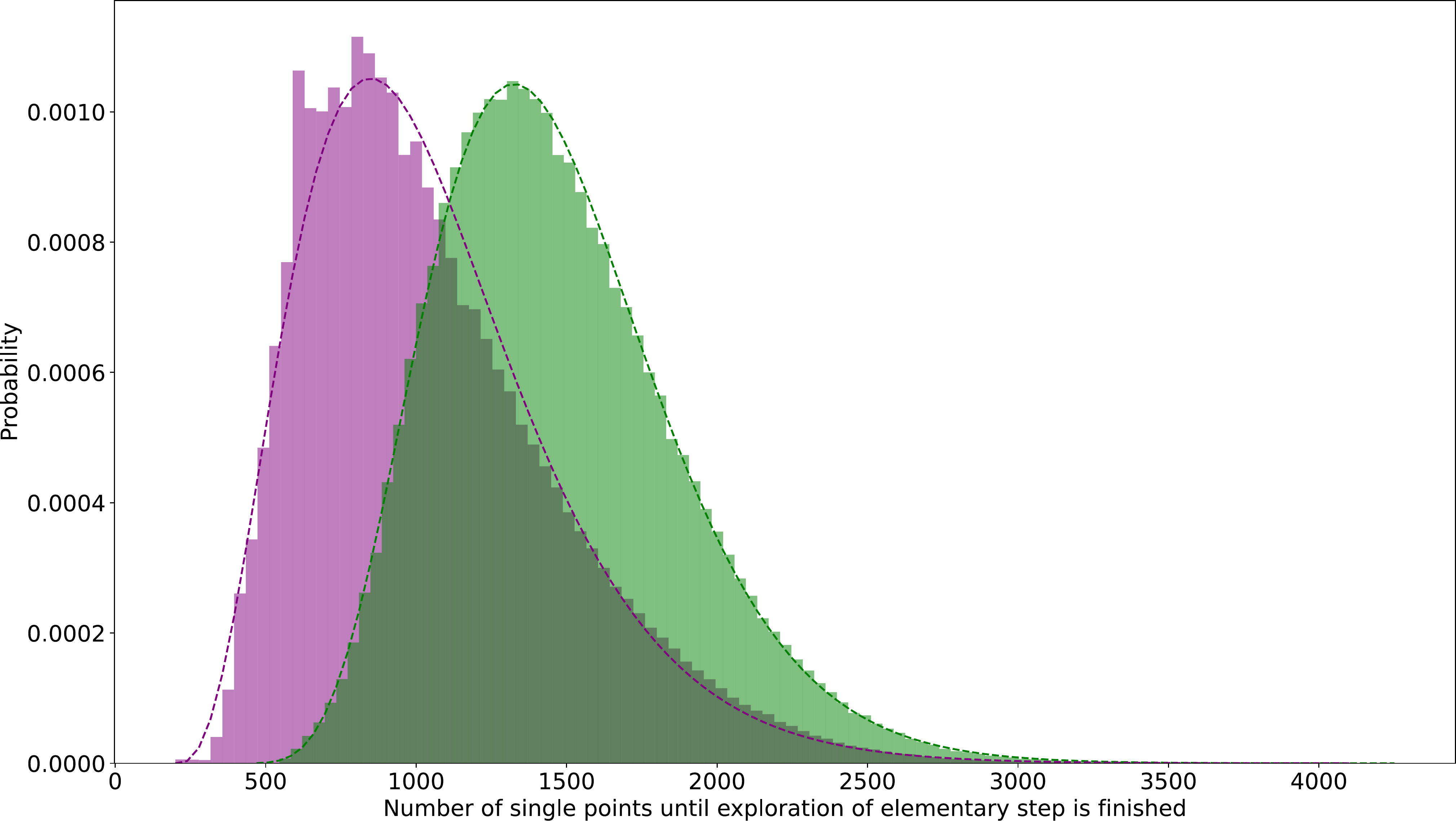}
   \end{center}
   \caption{\label{fig:sp-histogram}\small Histogram of the required number of single-point calculations for an elementary step search attempt (details see section~\ref{sec:computational_methodology}). Green bars represent successful and purple bars represent failed attempts.
   The dashed lines of the same color are the fitted $\gamma$ distribution.}
\end{figure}

The average number of single-point calculations per elementary step trial is depicted in a histogram in Fig.~\ref{fig:sp-histogram}, to which we fitted a $\gamma$-distribution due to the long tail towards higher numbers.
This fit allows us to estimate the number of calculations for successful elementary steps to be $1473 \pm 405$ and of failed attempts to be $1058 \pm 424$ (ranges defined by the standard deviation) for the current development version of our \texttt{Chemoton} software~\cite{Chemoton2021}.
However, a substantial number of unsuccessful attempts (11~\%) already failed within the first 200 steps of the Newton trajectory set-up because structures far away from an equilibrium structure were generated so that the self-consistent-field procedure did
not converge.
These calculations were excluded from the fit.
Upon taking them into account, the arithmetic mean of the single-point calculations required is lowered to 1050.

Structure optimizations of conformers generated with \texttt{Molassembler}~\cite{Molassembler2020,Sobez2020}
required only 3 days of total CPU time on a single core
for a total of $\approx 8.5 \cdot 10^6$ 
single-point calculations. Hence, it can be estimated that the costs of additional geometry optimizations, \textit{e.g.}, to refine structures based on more accurate electronic structure methods,
are negligible compared to elementary step trials.

\begin{figure}[H]
	\begin{center}
		\includegraphics[width=\linewidth]{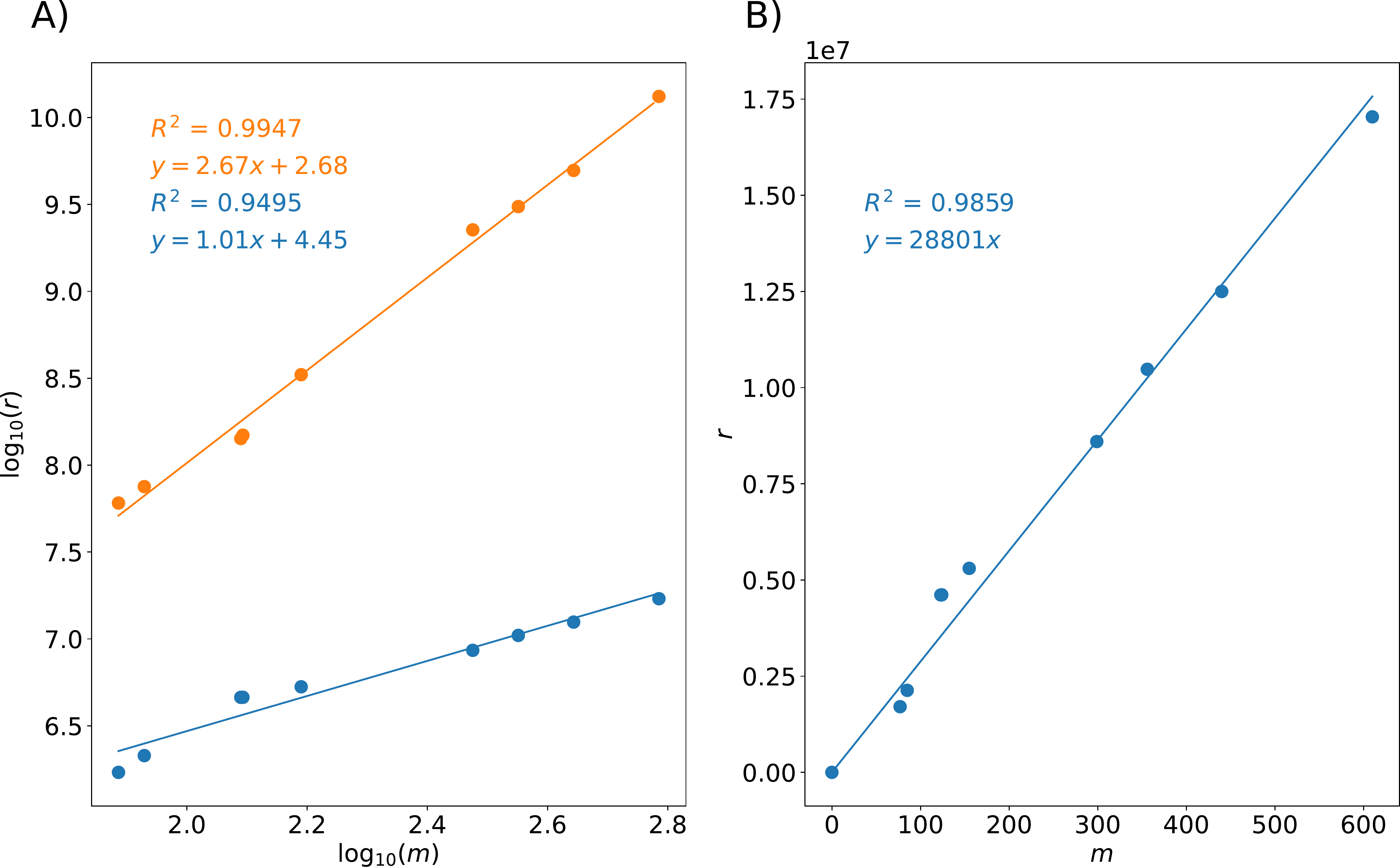}
   \end{center}
   \caption{\label{fig:scaling}\small \textbf{A)} A logarithmic plot of the number of elementary step trials $r$ against the number of compounds $m$ 
   in the reaction network within an upper limit for barrier heights of 200~kJ~mol$^{-1}$. The orange dots were calculated from eq.~(\ref{eq:scaling}) based on the number of compounds in the reaction network, while the blue dots are the actual number of elementary step trials based on the constraints applied during the exploration. \textbf{B)} Identical data points of the actual elementary step trials in the network, but without taking the logarithm.
   Lines represent linear regressions; the resulting linear equations are shown in the plot in the corresponding color.}
\end{figure}

Based on this extensive network, we can now study whether our assumptions about the scaling behavior 
were correct and how our graph distance restrictions
affect this scaling.
For this numerical analysis, we plot the number of elementary step trials $r$ against the logarithm of the number of compounds in the reaction network $m$ that are accessible within the given barrier height limit as shown in Fig.~\ref{fig:scaling}~A)).

It is evident that a quadratic scaling with the number of compounds
can be observed.
However, the total scaling is larger than quadratic, because the molecule sizes cannot be disregarded.
In addition, we understand that the chosen constraints based on the graph distance have a strong effect on the scaling behavior and reduce the scaling to a linear one.
Nevertheless, the slope of 28,000 of the linear scaling, shown in Fig.~\ref{fig:scaling} B), is still substantial, especially considering that we did not take into account the generated conformers in the reaction explorations, but probed possible elementary steps only for the first occurring conformer structure
of each compound.

Note also that explicit solvation was not considered in this extensive reference network.
Numerous approaches~\cite{Sunoj2012,JohnVarghese2019,Pliego2020,Simm2020,Steiner2021,Bensberg2021} 
exist that can limit the number of solvent molecules.
However, they still increase the required number of calculations and may require further development to tame this increased computational burden (\textit{e.g.}, by transferring solvation information with machine learning models from microsolvated nodes to those for which no microsolvation had been considered).

Whereas the network structure discussed so far did not contain any catalyst, we
now estimate how the addition of a homogeneous catalyst increases the computational resources required.
Formally, the scaling of the reaction network still follows the same pattern as before, because the catalyst molecule is simply another compound within the network.
However, because of the typical size of a catalyst of 50 to 150 atoms and because of the intricate relation between its structure and activity, the approximation that a single conformer is sufficient to provide a sufficiently deep and reliable overview on the reaction mechanism will, in general, no longer be valid.
Moreover, organometallic catalysts often represent challenging electronic structures, which can prohibit the application of fast semi-empirical methods, but require a more accurate description of the electronic wave function based, at least, on a fast (spin-unrestricted)
density functional approach.
Therefore, any practical exploration of a reaction network in the context of studying catalysis benefits from further restrictions in the exploration protocol, if they can be invoked without compromising the exploration depth.

\begin{figure}[H]
	\begin{center}
	   \includegraphics[width=0.8\linewidth]{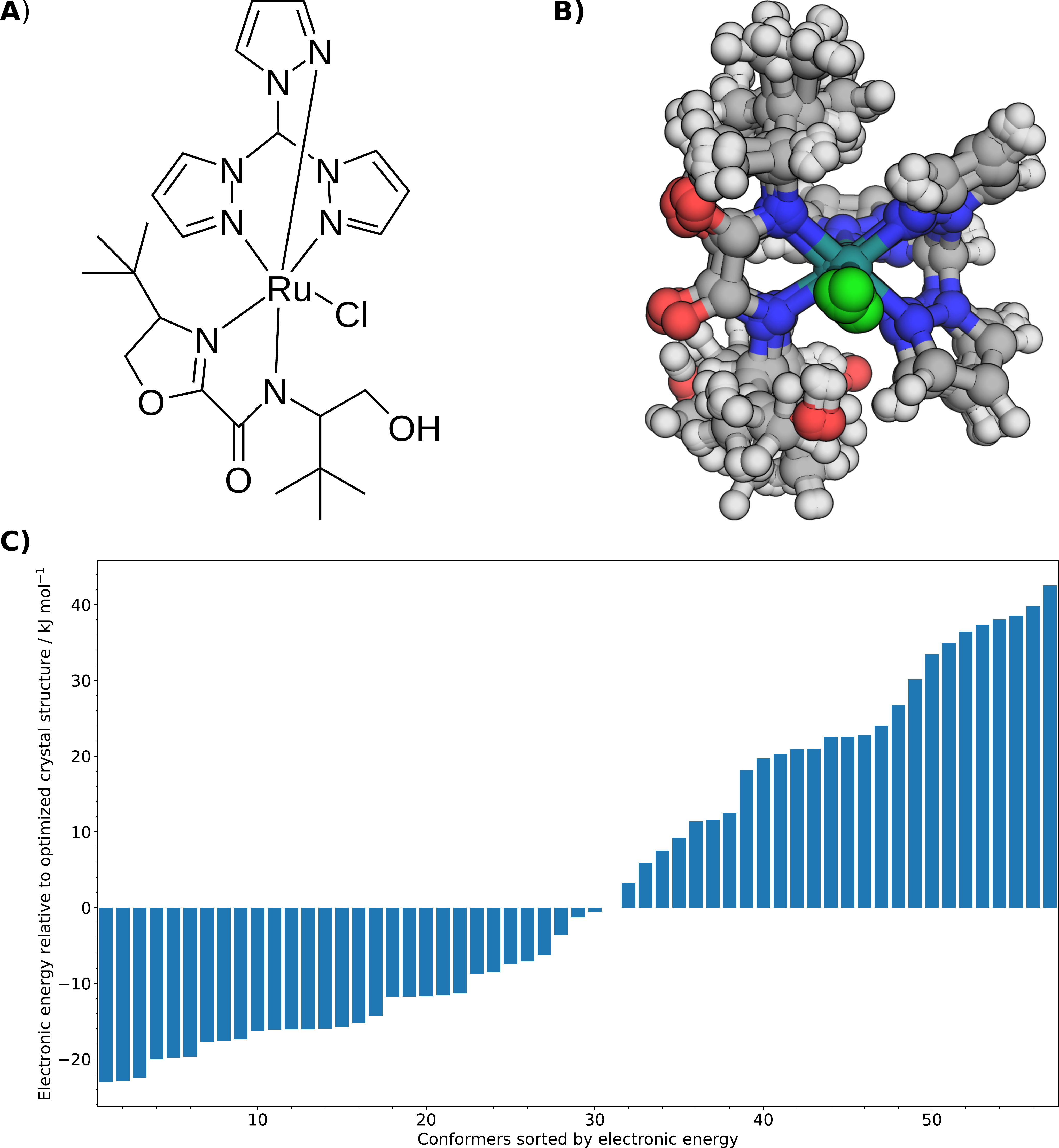}
   \end{center}
   \caption{\label{fig:conformers}\small Conformer analysis of an example catalyst, which we chose to be an organometallic catalyst for olefin epoxidation.
   \textbf{A)} Lewis structure of the catalyst;
   \textbf{B)} overlay of the optimized crystal structure and nine optimized conformers, which were the energetically lowest structures within their bin of structures after clustering;
    \textbf{C)} electronic energies of all 57 conformers relative to the optimized crystal structure.}
\end{figure}

Based on the data obtained for our reference network and a representative example of an organometallic catalyst, we now show, how severe such restrictions must be and how time-consuming exhaustive explorations of a catalytic reaction network can become.
We selected a ruthenium catalyst consisting of 66 atoms, which catalyzes the epoxidation of small cyclic olefins~\cite{Serrano2011}; see Fig.~\ref{fig:conformers}~A)).
We assume that a minimal catalytic cycle consists of around 10 different compounds and we can further estimate that the reaction mechanism including possible side reactions may be sufficiently well explored with 100 compounds, while 1,000 compounds would be a very exhaustive exploration of all reactions surrounding a catalytic cycle. Recall that
our definition of a compound~\cite{Unsleber2020} is a set of molecular structures with the same nuclear composition and connectivity; 
hence, one compound consists of numerous conformers.

Our uncatalyzed reaction network of 4,218 compounds starting from
propylene and oxygen already covers polymerizations, cyclizations, epoxidations, various peroxides, radical reactions, and beginnings of the formose reaction network.
To estimate the number of single point calculations $n_{sp}$ that are required to find $m$ different compounds 
we take the following metrics from our reference network and assume, as a starting point, that they are suitable for a network including a homogeneous catalyst:
\begin{itemize}
	\item success rate of elementary step trials $\sigma$
	\item ratio between elementary steps found and reactions $\varepsilon$, which yields an average number of elementary steps that belong to the same reaction
	\item average rate of newly found compounds per reaction $\eta$, as some reactions yield more than one previously unknown compound
	\item single-point calculations per elementary step trial $\nu$
\end{itemize}

Assuming that these metrics are independent of the number of compounds in the network, we arrive at eq.~(\ref{eq:compounds}) to
estimate the number of single-point calculation for constructing a network of $m$ compounds to be
\begin{align}\label{eq:compounds}
	n_{sp} &= \frac{\varepsilon \nu}{\sigma \eta} m \\
	       &\approx \frac{92.95 \cdot 1050}{0.22 \cdot 1.99} m \approx 2.2 \cdot 10^5 m.
\end{align}
However, all four parameters were taken from our reference reaction network and some of them will depend on the choice of our constraints in the exploration protocol.
For example, $\varepsilon$ will strongly depend on the number of conformers considered in the exploration and $\sigma$ can be increased with the application of a suitable reaction descriptor,
both of which were not considered in our reference network.
Therefore, we assume our $\varepsilon$ and $\sigma$ to be lower bounds for unguided explorations.

Based on these data, we can estimate the number of single-point calculations to find $10^2 - 10^3$ compounds to be approximately
$10^7 - 10^8$.
Any reactivity descriptors that identifies unreactive and reactive sites should manage to find all intermediates and products of the minimal catalytic cycle 
within these $10^2 - 10^3$ compounds, otherwise the number of required compounds and therefore calculations increases.

To estimate the number of conformers for an organometallic catalyst, we applied our conformer generation and optimization protocol implemented in \texttt{Chemoton} for our example catalyst. The crystal structure was taken from Ref.~\citenum{Serrano2011} and optimized with PBE-D3BJ/def2-SVP.
Our graph library \texttt{Molassembler} generated 57 conformer guesses, which were then optimized and the resulting structures were clustered according to root mean square deviation (RMSD) by \textit{average linkage agglomerative hierarchical} clustering 
with a distance threshold of 2.5~\AA (see 
the appendix), which resulted in nine representative conformers.
The results are shown in Fig.~\ref{fig:conformers}.

We expect a linear to quadratic effect of the number of considered conformers on the overall scaling, because conformers linearly increase the number of considered structures for explorations and in the worst case linearly increase the ratio of elementary steps and reactions $\varepsilon$ (assuming that all conformers still lead to the identical reaction).
Hence, the increase in the number of calculations for this example would be a factor of 100 in the worst case.
However, this would mean a consideration of about 10 conformers per compound in the network, 
which might not be necessary for most substrates.
Therefore, we may consider this number of conformers per compound as an upper bound requiring about $10^9 - 10^{10}$ calculations in a brute-force approach without the help of any pruning algorithms.
Based on the computing times for an energy and gradient of the crystal structure of the catalyst with the semi-empirical GFN2-xTB approach 
(\textit{i.e.}, 0.25 seconds per single-point in our set-up)
and with the generalized-gradient-approximation density functional with density fitting PBE-D3/def2-SVP (\textit{i.e.}, 2 minutes per single point in our set-up),
we extrapolate the required total CPU time to be $8-80$ and $4,000 - 40,000$ years, respectively.
In general, a reaction network exploration has the advantage of being trivially parallelizable, meaning that the use of $n$ computing cores brings an $n$-fold decrease in total wall time.
Therefore, the calculations for our example catalyst can be achieved with GFN2-xTB in $3-30$ days on 1,000 cores, while a complete exploration with DFT remains basically unfeasible without further modification of the exploration protocol or without a large increase in computing power.

In this context, it can be beneficial to carry out the time-demanding exploration trials with efficient semi-empirical methods and then refine the stationary points on a more accurate potential energy surface (PES).
In our reference network, the number of single-point calculations required for structure optimizations of stable intermediates
was three orders of magnitude smaller than the number of single-point calculations required for elementary step trials.
If we assume that $10^9 - 10^{10}$ single-point calculations are to be carried out for building a reaction network, we estimate another $10^6 - 10^{7}$ single-point calculations for a refinement of the network with a more accurate method, provided that the reaction mechanism or connectivity of the reaction network do not change significantly with the more accurate model.
Given our set-up for DFT calculations, this results in an estimate of $3 - 30$ years of computing time on a single core for the reaction network refinement, which again parallelizes trivially and could therefore be achieved within one day to two weeks on 1,000 cores. Note that this estimate will also be about the cost for every catalyst design feedback loop (discussed in section~\ref{sec:concept:design}) if the design shall be based on rigorous first-principles-based reaction network information.

These estimates do not consider any restriction or constraint in the exploration process itself. 
Apart from the pruning options already discussed above (i.e., first-principles heuristics for reactivity descriptors \cite{Bergeler2015,Grimmel2019,Grimmel2021}
and exclusion of short-lived intermediates from further exploration), 
the exploration process may be kinetically driven by steering trial and search calculations to those parts of the network that can be reached under reaction conditions by exploiting barrier information~\cite{Simm2017} or explicit micro-kinetic modeling~\cite{Proppe2019}. 
Hence, we may assume that broad automated reaction network explorations are within reach, provided that reliable approximate methods are available and the exploration space can be limited without excluding important reactions.

Unfortunately, resource estimates for explorations of heterogeneous catalysts cannot easily be inferred from data on homogeneous systems.
For heterogeneous catalysis, we need to consider additional structures and elementary steps to bridge the phase difference between catalyst and reactant as discussed in the next section.

\subsection{Special algorithms for heterogeneous catalysis}
\label{sec:hetero_algorithms}
Typical heterogeneous catalysts exhibit vastly different structural motifs compared to molecules in the gas phase or in solution, which need to be accounted for in the exploration.
The algorithms that we implemented in \texttt{Chemoton} for this work in order to resolve these
challenges are described in this section.

Any extensive exploration requires to compare individual structures in a timely manner. Root mean square deviations of Cartesian coordinates are not suitable for the process for various reasons (\textit{e.g.}, they depend on molecular size and will require elaborate thresholding for making reliable statements on molecular identity).
Graphs are among the best options for such a metric, because (i) they can be compared efficiently and do not depend on system size, (ii) they are chemically intuitive, and (iii) they allow for substructure/similarity searches.
In the automated explorations conducted so far, we exploited graph-based comparisons that are facilitated by the \texttt{Molassembler} library~\cite{Molassembler2020,Sobez2020}.
To construct graphs, connectivity information is required, which may be taken from simple distance information or from population analysis of electronic wave functions that yields quantum chemical bond order information. For solid-state systems such as those acting as catalyst or catalyst supports in heterogeneous catalysis, this information is not straightforward to obtain (\textit{e.g.}, consider the adsorption process and how an adsorbate's binding to a surface is to be characterized in terms of chemical bonding).

The seemingly easiest approach to determine bonds in a three-dimensional structure is distance criteria.
Parametrized distances for each element are sufficient for molecular structure, but often fail for solid state structures.
The two remaining distance-based approaches are Voronoi tessellation and nearest-neighbor criteria.
Voronoi tessellation fails for surface systems without the knowledge of the corresponding crystal structure~\cite{Boes2019}; hence, it is difficult to implement within an automated exploration algorithm, where each minimum structure has to be labeled with a graph, which should ideally only be dependent on the structure's spatial coordinates and electronic structure and not be based on inheritance from other structures.

Nearest-neighbor approaches work well for crystal and surface structures, 
but can fail for molecular structures, because the atoms in molecules have varying elements as bonding partners with different bond lengths. Therefore, an approach to detect bonds only between the closest distances would either overlook valid bonds or 
require an elaborate inclusion threshold.
Hence, an algorithm solely based on distances must know which nuclei are part of a solid state structure and which are part of an adsorbate.
Additionally, the algorithm must then select the distance criterion based on this categorization of nuclei within one structure
and also be able to handle chemical and physical adsorption.
Such elaborate tracking of nuclei and categorization
can introduce many system-dependent heuristics and possible points of failure within an automated exploration.

Alternatively, bonds may better be derived directly from the electronic structure, which avoids system-dependent heuristics.
We implemented Mayer bond orders~\cite{Mayer1983} in \texttt{SCINE} 
for molecular and periodic structures, which allows us to directly compare the different approaches.
Alternatively, DDEC6 bond orders based on a so-called \textit{dressed exchange hole} determined by the electron density distribution, which has been tested for a wide array of chemical structures
~\cite{Manz2017}, may provide more reliable bond estimates.

Adsorption is a key feature of heterogeneous catalysis that is absent in homogeneous catalysis.
However, a selection of every nucleus and bond as a potential active site would make an automated exploration unfeasible.
In some cases, active sites are likely to be found on high symmetry sites of the surface~\cite{Ertl2008}.
Accordingly, Persson~\textit{et al.}~applied a Delaunay triangulation on the top layer of the surface slab to retrieve \textit{top}, \textit{bridge}, and \textit{hollow} sites from the corners, edges, and centers of the triangles~\cite{Montoya2017,Andriuc2021}.
The number of these sites can then be reduced based on the symmetry of the surface structure.
Boes \textit{et al.}~\cite{Boes2019} improved the algorithm by first constructing a graph of the corresponding crystal structure with Voronoi tessellation.
This allows one to identify the top layer nuclei of any surface resulting from that crystal structure and to construct an adsorption direction based on the normal vector of a plane spanned by all neighboring atoms in the surface graph.
Deshpande \textit{et al.}~\cite{Deshpande2020} directly inferred the adsorption sites from the surface graph, but constructed the graph
with a nearest-neighbors approach. 
This procedure allowed them to deduplicate the relaxed surface structures according to their local-graph information.
Recently, Marti \textit{et al.}~\cite{Marti2021} released the software \textit{DockOnSurf}, which was specifically developed to generate structures for complex adsorbates
and surfaces based on pre-screening of conformers, adsorbing them based on geometric centers of nuclei, and screening conformers on the surface according to dihedral angles.
The resulting structures were then deduplicated following an energy criterion.

\begin{figure}[H]
 \begin{center}
    \includegraphics[width=0.8\linewidth]{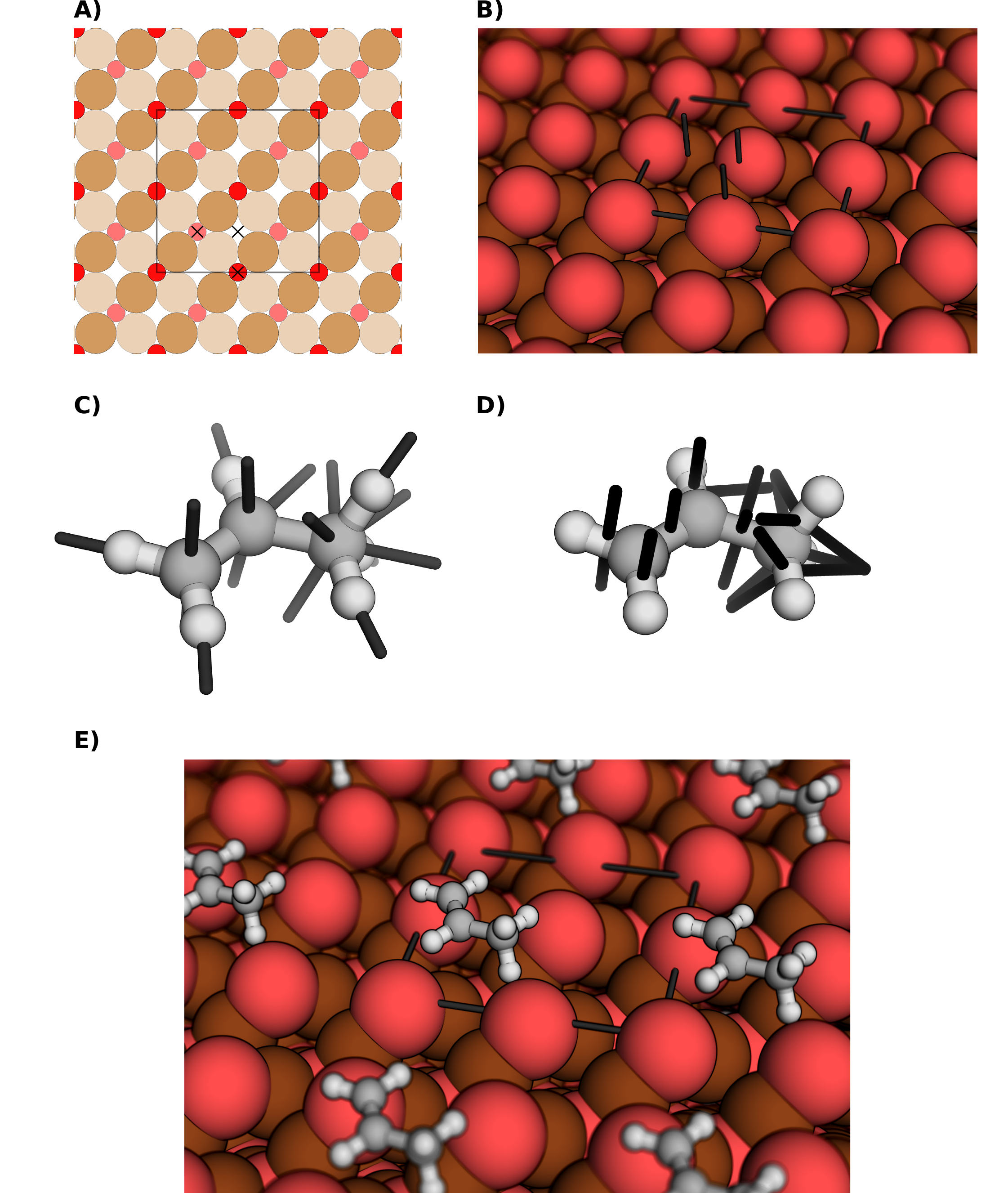}
\end{center}
\caption{\label{fig:adsorption}\small Representation of the adsorption workflow implemented in \texttt{Chemoton}: A) two-dimensional view of a Cu$_2$O~(001) slab with detected adsorption sites marked by black crosses and the unit cell by black lines; B) three-dimensional view of the slab with the unit cell and the adsorption vectors marked by black sticks; C) directions of attack indicated by black sticks for each nucleus in propylene and in D) for each bond in propylene; E) example for an adsorbed structure after structure optimization with PBE-D3BJ/DZVP-MOLOPT-GTH.
}
\end{figure}

For this work, we adopted the already existing general algorithms
in \texttt{Chemoton}, which were developed for intermolecular reactions~\cite{Chemoton2021},
to establish a new adsorption algorithm which can handle surface slabs and nanoparticles, multidentate adsorption and any adsorbate, while also minimizing the number of screened structures.
This new workflow is illustrated in Fig.~\ref{fig:adsorption}.

In the case of surface slabs, \texttt{Chemoton} first detects the high symmetry sites based on Delaunay triangulation as shown in Fig.~\ref{fig:adsorption}~A) and implemented in \textit{pymatgen}~\cite{Ong2013}. Then, \texttt{Chemoton} determines an adsorption vector based on steric hindrance, which allows the program to determine the optimal angle of adsorption, while requiring no graph information of the surface structure.
The vectors corresponding to the detected sites are illustrated in Fig.~\ref{fig:adsorption}~B).
The adsorbate is then treated as an intermolecular
reaction partner and the directions of attack can be formulated for any combination of active sites within the molecule as shown in Fig.~\ref{fig:adsorption}~C) for nuclei and in D) for bonds, which can be extended to any complex combination of multiple nuclei.

The adsorption guess structure is then simply generated by alignment of the direction vectors and can additionally be diversified by considering multiple rotamers defined by a rotation around the direction vectors.
The generated guess structure can be optimized with any of the available quantum chemistry programs within \texttt{SCINE} (see appendix)
and an example result is shown in E).
This workflow allows us to reduce the number of explored structures based on symmetry while also being able to treat any chemical system.
If no significant symmetry is present, \textit{e.g.,} for nanoparticles, we apply the standard intermolecular approach implemented in \texttt{Chemoton}.

For systematic autonomous explorations, the generation of multiple adsorption structures of a single compound
is not sufficient, but requires two more steps.
First, the number of subsequently explored structures must be reduced by deduplication analysis after structure optimization, because different guess structure may lead to the same minimum.
Since an energy criterion for deduplication does not directly relate to structural equality, it may lead to false positives and may hide crucial branches of the reaction network, we opt for a graph-based approach, which is required for large scale explorations in any case.
Second, a first-principles-based exploration requires to sample different reactions, which, in the context of heterogeneous catalysis, often requires to adsorb multiple different reactants onto the same surface slab.
This is an algorithmic problem, which has hardly been discussed in the literature.

The existence of an already adsorbed molecule causes three main issues in the context of automated adsorption protocols.
First, the algorithm must be able to distinguish the existing adsorbate from the remaining surface slab, otherwise it would be
detected as a surface site, which may lead to the generation of inaccessible high-energy structures.
Second, the existing adsorbate breaks the symmetry of the surface
slab in most cases and the number of different second adsorption structures is therefore significantly larger.
Finally, the surface may have changed after the first adsorption step, which may prohibit to infer second adsorption positions from the structure of a clean slab.

Therefore, we implemented an algorithm within our automated exploration that tracks which nuclei are part of the surface and which are not. It is able to execute a modified Delaunay triangulation without symmetry exclusion, but with steric exclusion of sites too close to the first adsorbate.
This leads to a plethora of possible sites, especially for larger slab models.
Hence, it is often wanted to minimize the second adsorption step to sites that are within a reasonable distance to the first adsorbate, especially since the exploration should sample
potential reactions of the adsorbed molecules.
Additionally, the exploration should also consider that the second molecule may directly react with the adsorbate from the gas phase,
which is why one must screen for such possibilities.

The adsorption algorithm discussed here now allows us to generalize our resource requirements analysis from a homogeneous reaction network to a heterogeneous one.

\subsection{Resource estimates for automated exploration of heterogeneous catalysts}
\label{sec:ressource_estimates_hetero}

Because a heterogeneous catalyst is \textit{per se} only another compound in the network, we know that our reference reaction network consisting of molecules only would be formed identically if we did not enforce any limitations or favored heterogeneous reactions.
As in the case of homogeneous catalysis, we cannot consider a single structure of the catalyst only.
However, the definition of 'conformers' is, of course, very different for solid state structures, which we discuss in
the following.
We will also see that new definitions for elementary steps are required which are elaborated on afterwards.

Conformers of a heterogeneous catalyst are not necessarily formed from already active structures, but rather stem directly from the crystal structure.
For regular surfaces, these are usually discussed in terms of their Miller indices, including defects and different terminations of the surface.
A consideration of all possible surfaces is impossible due to their infinite number and the consideration of many,
\textit{e.g.,} $\ge 10$, is hardly considered in manually
guided studies, which can afford only fewer calculations per discovered compound and need
to exploit preexisting knowledge.

For well characterizable surfaces, we may roughly categorize automated heterogeneous explorations in terms of the number of the surfaces (plus decoration) considered per solid state catalyst.
A minimal exploration would consider only a single surface without defects. 
An extensive exploration would consider the (100), (110), and (111) surfaces, usually termed \textit{low-index surfaces}, with different surface terminations, as clean surfaces and a point vacancy and adatom each to include effects of the most-common defects.
An exhaustive exploration would consider every surface up to a maximum Miller index of four ($\approx 30$ surfaces), every possible surface termination as clean surfaces and with $\approx 5$ different defects each.
Before estimating the scaling of the number of surfaces, we first introduce the term of the number of unique elemental species $e$.
This shall be defined as the number of types of atoms existing in a solid state structure, if all atoms are categorized based on their element, local coordination, and electronic properties.
We can roughly estimate that $e$ linearly increases the number of possible surface terminations and possible point defects each.
The number of surfaces to be considered, $n_\text{surf}$, is then given by
\begin{equation}\label{eq:n_surfaces}
	n_{\text{surf}} = n_{\text{indices}} \cdot (n_{\text{termination}}(e) \cdot n_{\text{defects}} \cdot e + n_{\text{termination}} (e)).
\end{equation}
For a bielemental crystal and $n_\text{termination} (e) \approx e \approx 2$, we estimate $n_{\text{surf}}$ in the three exploration protocols termed above as 'minimal', 'extensive', and 'exhaustive' to be 2, 30, and 600, whereas $e=3$ would increase $n_{\text{surf}}$ to 3, 63, and 1500.
Of course, the number of considered Miller indices $n_{\text{indices}}$, surface terminations $n_{\text{termination}}$, and defects $n_{\text{defects}}$ are completely independent of each other and explorations can be envisioned that only focus on one of these aspects to decrease the computational costs.

\begin{figure}[H]
	\begin{center}
	   \includegraphics[width=0.7\linewidth]{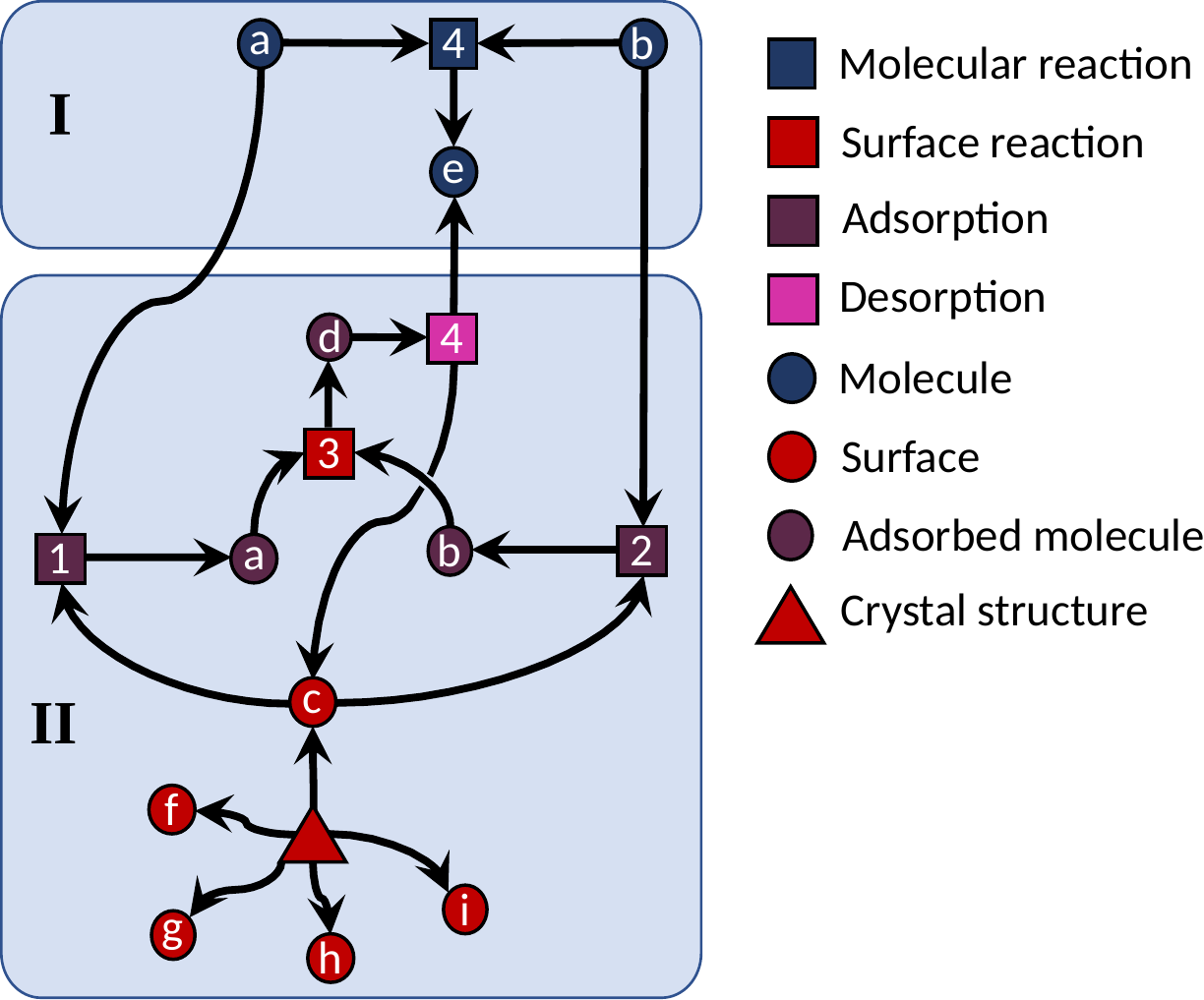}
   \end{center}
   \caption{\label{fig:hetero-network}\small A minimal reaction network is shown consisting of a molecular part \textbf{\MakeUppercase{\romannumeral 1}} and a solid state interaction part \textbf{\MakeUppercase{\romannumeral 2}}. It includes two compounds $a$ and $b$, and one surface $c$. The two compounds can react to $e$ uncatalyzed via the blue reaction 4 or catalyzed by $c$ through the series of reactions 1-4.}
\end{figure}

For a given number of surfaces considered, $n_{\text{surf}}$, which do not mutually affect the exploration of one another, we can estimate the scaling of the elementary steps for each of them
so that the total scaling will be linear in $n_{\text{surf}}$.
While the purely molecular part of the exploration is not changed by the addition of a heterogeneous compound, new types of elementary step trials $r_{\text{surf}}$, which scale differently when compared to purely molecular elementary step trials $r_m$, must be introduced into the network exploration.
Furthermore, the number of possible compounds varies for this part of the network, which is why we split the total number of compounds $m$ into molecular compounds $m_m$ and compounds adsorbed on surfaces $m_s$ for our scaling estimates.
Moreover, we can split any additional elementary step trials involving the solid phase into adsorption trials $r_a$, trials between surface species $r_s$, and desorption trials $r_d$, which yields the total number of elementary step trials $r$ as
\begin{equation}
	r = r_m + n_{\text{surf}} \cdot r_{\text{surf}} = r_m + n_{\text{surf}} \cdot \left( r_a + r_s + r_d \right).
\end{equation}
The scaling of $r_m$ was already evaluated and discussed in section~\ref{sec:ressource_estimates_homo} and shown in eq.~(\ref{eq:scaling}).
The elementary step trials for adsorption, $r_a$, can be considered as special cases of intermolecular reactions with identical scaling to eq.~(\ref{eq:scaling}) for the molecules, 
whereas each considered surface is only a multiplicative value based on its available first adsorption sites $n_{\text{sites}_1}$, which gives $r_a$ as
\begin{equation}\label{eq:adsorb_trials}
	r_a = n_{\text{sites}_1} \sum_{i=1}^{m_m} \sum_{a=1}^{n_i} n_{ci} d_a \rho_a {n_i\choose a}
\end{equation}
with $n_{ci}$ for the number of conformers of compound $i$ considered for elementary steps.
Similarly, the trials for elementary steps on surfaces $r_s$ can also be considered as intermolecular reactions with the number of second adsorption sites $n_{\text{sites}_2}$ in place of the different directions of attack $d_a$ and rotamers $\rho_a$, which leads to
\begin{equation}\label{eq:surface_trials}
	r_s = \sum_{i=1}^{m_s} \sum_{j=i}^{m_s} \sum_{a=1}^{n_i} \sum_{b=1}^{n_j} n_{\text{sites}_2} n_{ci} n_{cj} {n_i\choose a}{n_j\choose b}.
\end{equation}
In a brute-force approach, every adsorbed compound must also be probed to be desorbed or dissociated.
The number of elementary step trials for complete dissociation of an adsorbed compound is equal to the number of compounds, while dissociations of adsorbed compounds can be viewed as intramolecular dissociations, which gives 
\begin{equation}\label{eq:desorb_trials}
	r_d =  \sum_{i=1}^{m_s} n_{ci} + \sum_{a=1}^{\left \lfloor{\frac{n_i}{2}}\right \rfloor} \sum_{b=a}^{n_i-a} n_{ci} {n_i\choose a+b}.
\end{equation}
If we now apply identical constraints on the number of possible active sites in a molecule as in our reference network (such as limiting intramolecular dissociation trials to repulsion of bonded nuclei or maximally combining bonds and bonds in intermolecular reaction trials), 
we arrive at
\begin{align}
	r_{surf} &= \sum_{i=1}^{m_m} \sum_{a=1}^{2} n_{\text{sites}_1} n_{ci} d_a \rho_a {n_i\choose a} + \sum_{i=1}^{m_s} \sum_{j=i}^{m_s} \sum_{a=1}^{2} \sum_{b=1}^{2} n_{\text{sites}_2} n_{ci}  n_{cj} {n_i\choose a}{n_j\choose b} \\
	&+ \sum_{i=1}^{m_s} n_{ci} +  n_{ci} \frac{n_i^2 - n_i}{2}.\nonumber
\end{align}
These additional types of reactions, which are typical for reactions on surfaces, are highlighted in a minimal reaction network in Fig.~\ref{fig:hetero-network}.
In that figure, the molecular reaction network \textbf{\MakeUppercase{\romannumeral 1}} in blue is enhanced by the reaction network \textbf{\MakeUppercase{\romannumeral 2}}, which consists of interactions with solid state structures.

In Fig.~\ref{fig:hetero-network}, $r_a$ corresponds to reactions 1 and 2, reaction 3 resembles $r_s$, the pink reaction 4 (as well as the reverse reactions of 1 and 2) corresponds to $r_d$, and the blue reaction 4 is an example for $r_m$ and is in general the uncatalyzed variant of the series of reactions shown in network \textbf{\MakeUppercase{\romannumeral 2}}.
In general, $r_a$ scales quadratically with $n_i$, linearly with $m_m$, and linearly with the number of possible first adsorption sites $n_{\text{sites}_1}$.
However, the scaling with $n_i$ can again be reduced by exploiting graph constraints as shown in section~\ref{sec:ressource_estimates_homo}.
We may also assume that $r_s$ scales similar to our reference network, although the slope of the linear scaling might be larger due to the factor of $n_{\text{sites}_2}$. 
\textit{A priori} $n_{\text{sites}_2}$ is considerably larger than $n_{\text{sites}_1}$ due to missing symmetry as discussed in section~\ref{sec:hetero_algorithms}. However, close-proximity constraints can limit this to an approximately constant number of sites on the order of 10.
By contrast, $n_{\text{sites}_1}$ depends on the complexity of the surface slab and it can be estimated to scale linearly with the number of unique surface species $e$.

The exact scaling of the elementary step trials $r_s$ and $r_d$ is difficult to estimate, because they only apply for compounds that include adsorbed species.
As shown in eq.~(\ref{eq:m_s}), $m_s$ depends on the success rate $\sigma$ of the adsorption elementary steps $r_a$ and the number of elementary steps $\varepsilon$ that are found for the same reaction 
\begin{equation}\label{eq:m_s}
	m_s = \frac{\sigma}{\varepsilon} r_a.
\end{equation}
However, the assumption that $\sigma$ and $\varepsilon$ are similar in value compared to our uncatalyzed molecular reaction network is not valid.
While the screening algorithms within \texttt{Chemoton} are very similar, the underlying chemical processes are too different to expect similar numbers and they will, in general, also vary between different surfaces.
Due to this dependence on the chemical structure, we cannot provide valid general estimates of the number of elementary steps and therefore on the number of single-point calculations required for a heterogeneous network.
However, based on the fact that a single surface does not require conformer generation and its different adsorption sites can be viewed as similar to directions of attack in a molecular structure (albeit with a slightly different scaling), the order of magnitude of required single-point calculations for a purely heterogeneous network should be similar to a homogeneous one.

The largest cost factor is, instead, the number of considered surfaces $n_{\text{surf}}$, which linearly increases the number of all calculations.
Therefore, the computational costs of an exploration would be increased by a factor of 1,000, if various surfaces and defects would be considered as shown earlier in this section.
Although it can be assumed that 100 would already cover most relevant reactions and 10 may be enough based on restrictions that may be deduced from experimental data.
If we therefore assume a factor of 100, the required number of single-point calculations in a brute-force approach may be similar to those required in the exploration of a homogeneous catalyst, which we deduced to be $10^9 - 10^{10}$.

Due to the inherently larger number of atoms in solid state structures (and imposed periodic boundary conditions), the calculation times are usually longer compared to molecular systems.
If we again take the example of propylene epoxidation, for which Cu$_2$O is a potential catalyst~\cite{Khatib2015,Duzenli2015,Porter2021}, we can estimate the total computing time based on the time required for a single calculation of a (001) slab with an extension of 2$\times$2$\times$3, 
which can be taken as the minimal slab size for the exploration of such a reaction.
This then leads to a total computing time of $10^6 - 10^7$ years on a single core.

\subsection{Requirements for predictive computational catalysis}
\label{sec:concept:predictive}

To make reliable and accurate \textit{in silico} predictions about catalytic processes, the following requirements need to be fulfilled.

First, it must be guaranteed that all accessible reactions under some specified ambient conditions are explored. Since there is no way to know that everything has been found in an exploration, this can never be guaranteed. However, it needs to be shown in computer experiments that the exploration algorithms chosen can reproduce the relevant parts of a reference network. Clearly, such reference networks for diverse catalytic systems must first be developed, which will require a community effort. While the heuristic nature of this approach cannot be circumvented, it is clear that the exploration algorithms must be general (\textit{i.e.}, agnostic with respect to all sorts of chemical constraints) and cover all relevant reaction types.

Second, the uncertainty of predictions must be accessible, which will require error estimates for all key quantities in the exploration process.
Since it is impossible to derive accurate errors for many-particle problems in quantum mechanics (otherwise, an accurate 
quantum mechanical solution would have been found and the approximations would no longer be needed, which is impossible for any relevant catalytic system), a Bayesian approach is required that transfers error estimates obtained for some nodes after investment of additional computational resources to nodes for which such information is not available~\cite{Proppe2017,FaradayUncertainty2017,Simm2018,Proppe2019}.

Third, structural fidelity, \textit{i.e.}, the fact that the nuclear scaffolds that define the external potential in the quantum chemical calculations sufficiently well represent the chemical system in terms of molecular structure, surface, and solvent, needs to be ensured for all predictions.
Only if the structural model adequately resembles the experimental situation, reliable predictions can be made.

Finally, it should be possible to use electronic structure methods applied interchangeably in order to find the best compromise between accuracy and speed by switching from fast-approximate to expensive-accurate methods.
Such switches can either be driven in an automated fashion, if a suitable descriptor (such as confidence intervals from machine learning models~\cite{Simm2018}) is available, or the software issues a warning and requires manual intervention~\cite{Simm2019}.
These approaches must be combined into general workflows, some of which will be discussed in the following section.

We emphasize that the diversity of all reaction steps that can occur is so vast, even if one restricts the exploration to the known 
ingredients (\textit{i.e.}, ignoring the unknown ones such as impurities in solution or at a surface), that achieving completeness is formally impossible. This is not a key problem of an autonomous approach that targets orders of magnitudes more detail (measured, \textit{e.g.},
in terms of the number of elementary steps or the number of potentially important impurities such as traces of oxygen or water
in a reaction liquor) than what could be inspected manually. However, manual intervention is, of course, possible and can be
used to steer an exploration into specific regions of chemical reaction space by letting the search algorithms probe reactants
that are potentially and unintentionally present in the experiment. It is for this reason that we have begun
to estabish interactive quantum mechanics 
\cite{Haag2011a,Haag2014a,Haag2014,Muhlbach2016a,Vaucher2016,Vaucher2016b,Heuer2018,Vaucher2018}
for an easy and simple interference of an operator with an autonomously running exploration protocol.

\subsection{Workflows for efficient computation protocols}
\label{sec:concept:worflow}

As shown in sections~\ref{sec:ressource_estimates_homo} and \ref{sec:ressource_estimates_hetero}, even if a single calculation may be efficient, the amount of data generated in an exhaustive exploration is immense and on the scale of $10^9$ single-point calculations in brute-force approaches.
Therefore, smart automated protocols must be established to steer the exploration and reduce the number of calculations in order to maintain efficiency through all stages of the exploration process.
A general paradigm for these workflows should be the automated selection of the minimally required algorithm for each specific task, while still being transparent, so that the applied approximations and their limitations can be understood.
This requirement inherently requires flexible and modular workflows.

A prime challenge, which demands such an approach, is conformer generation.
The generation of conformers of a chemical structure is necessary to reflect the structural ensemble accessible at a given temperature.
This has been of major importance in the design of new pharmaceuticals~\cite{Hawkins2017}; hence, most conformer generation algorithms have been tested and compared on drug-like molecules~\cite{Ebejer2012,Friedrich2017}.
However, the importance of conformers in the elucidation of reaction mechanisms and the calculation of reaction barriers has also been emphasized recently~\cite{Simm2017,Vitek2020,Viegas2021}.

The most efficient methods for sampling the phase space of a chemical structure apply prior chemical knowledge to systematically generate conformers with rotations around rotatable bonds.
Such algorithms can be developed based on heuristic rules~\cite{Leite2007,Miteva2010,Hawkins2010,OBoyle2011a,Poli2018,Gavane2019,Friedrich2019},
distance geometry~\cite{Vainio2007,Riniker2015,Sobez2020},
machine learning~\cite{Gebauer2018,Mansimov2019,Chan2019,Chan2020,Gogineni2020,Simm2020a,Fang2021,Ganea2021},
or methods beneficial for quantum computing~\cite{Marchand2019}. 
However, due to the combinatorial increase of possible conformers with the number of rotatable bonds, all these algorithms become unfeasible at a certain system size and stochastic sampling will be required. Hence,
at this point, the conformer generation method must switch to algorithms that do not aim at covering the complete phase space, but sample most relevant regions of the PES within reasonable time.
The most common examples in this regard are MD simulations with enhanced sampling techniques.
Due to the plethora of different enhanced sampling techniques developed, we refer the reader to recent reviews~\cite{Abrams2014, Bernardi2015, Tiwary2016, Yang2019} for discussions of their differences and advantages,
and to Refs.~\citenum{Kamenik2018,Zivanovic2020,Pracht2020} for different applications in the context of conformational sampling.

Since MD simulations are inherently expensive in terms of computing time, it is beneficial to additionally apply a multilevel approach for the evaluation of the PES.
Larger systems can first be evaluated with faster, less accurate models and the most relevant conformers can later be studied with more accurate methods~\cite{Chandramouli2019,Grimme2021}.
However, within some finite computing time given also these algorithms will eventually fail for increasingly larger systems. 
The situation will then be similar to that of the prediction of a most stable protein fold, which is a conformational sampling problem at its core and for which specific knowledge-based approaches are advantageous~\cite{Senior2020,Baek2021}.

Should the end of a chain of available algorithmic switches be reached, the (meta)algorithm that implements the 
switching must recognize that the problem cannot be solved in an autonomous fashion. Structural fidelity can no longer
be guarantedd and manual intervention might be required; \textit{e.g.}, the algorithm may warn the operator as already discussed in Ref.~\cite{Haag2014a}.
Such cases could then be approached within an interactive setting~\cite{Haag2014a,Haag2014,Simm2017,OConnor2018}.

Another well-known problem, which requires a sequential switching approach with maximum automation and minimal, but intuitive interaction, is the search for transition states (TS), \textit{i.e.,} first-order saddle points on a PES.
Numerous stable and reliable TS optimization algorithms have been developed in the last fifty years~\cite{Schlegel2011,Henkelman2017,Bofill2020}.
However, due to the difficult nature of the optimization problem, a universally successful algorithm that is able to find all relevant TS from a limited number of start conformations will most likely never exist.
Therefore, the software must be able to recognize that two or more structures should be connected via a TS, although a series of attempts of various algorithms has failed to locate a TS.
This recognition can be based on physical or structural descriptors such as RMSD or graph comparisons.
In such a case, the software can present the issue to the operator in an interactive manner, who can decide, whether this possible reaction is relevant and may even provide another educated guess for the TS based on real-time quantum chemistry~\cite{Haag2014,Heuer2018,Vaucher2018}.

Finally, we discuss required workflows to model heterogeneous processes based on the algorithms outlined in section~\ref{sec:hetero_algorithms}.
In general, heterogeneous reactions can be explored with two different approaches.
On the one hand, the chemical reactions can first be explored for molecules in the gas phase and in a second step all possible intermediates can be transferred onto the heterogeneous catalyst in an adsorption step.
This saves computing time by minimizing the exploration trials in the solid state, which requires longer computing times, and enables double-ended searches for the reactions in the adsorbed state.
However, this approach may fail if the intermediates of the heterogeneously catalyzed reaction are significantly different to those in the gas phase.

On the other hand, the exploration can proceed by screening for potential reactions directly in the solid state, which was outlined in our resource estimates in section~\ref{sec:ressource_estimates_hetero}.
In this approach, the reactants are adsorbed on a minimal surface slab directly and screened for conformations, either on the surface directly or by adsorbing various conformers.
Then, the ranked adsorbed conformations of multiple reactants can be combined on a surface slab, which may require an extension of the surface.
There, the second adsorption sites must be limited based on distance constraints and preferable adsorption sites already screened in the first adsorption as discussed in section~\ref{sec:hetero_algorithms}.
The various possible extensions of the solid state structure must also be carefully stored and evaluated in reaction energy analysis across the reaction network.

\section{Conclusions}
\label{sec:conclusion}

Autonomous reaction network exploration presents an innovative, unbiased, and expansive approach of studying chemical reactivity.
In this work, we discussed the potential of understanding catalytic processes in terms of
automatically generated reaction networks from first-principles calculations and elaborated on required concepts and workflows.
First-principles-based approaches are expensive in terms of computer time, but they are indispensable if detailed
mechanistic insight is sought for. High-throughput experimentation and data mining are complementary and may even
deliver results for catalyst design purposes much faster than first-principles
calculations. However, first-principles modeling is also appropriate in cases where experiments are difficult to conduct
(\textit{e.g.}, in high-throughput settings) or where data are incomplete. 

As an example, we estimated the computational costs associated with exhaustive first-principles explorations in brute-force approaches for
a reference reaction network of $10^6$ structures constructed by starting with two reactants.
Our resource estimates showed that truly extensive explorations based on density functional theory calculations
without activated pruning schemes (to cut deadwood in the exploration) are not feasible 
because of the sheer number of exploratory calculations to be carried out.
This can be alleviated by suitable first-principles reactivity descriptors \cite{Bergeler2015,Simm2017,Grimmel2019,Grimmel2021} which not only
can suggest potentially reactive sites to be prioritized in the exploration process, but which can also determine
those sites that are likely to be unreactive and that can therefore be given a very low priority in the exploration process.

The efficiency of building reaction networks with time-independent calculations is also increased
by exploiting the fact that it parallelizes in a trivial manner because many elementary reaction trials can be carried out
in parallel. 
Moreover, fast-but-very-approximate semi-empirical calculations 
can be employed for acquiring quickly a broad overview on a network. 
The key property of a suitable semi-empirical method must be structural fidelity since an energy refinement
can be done in a subsequent step. In an autonomous setting, this is most efficiently accomplished by automated determination 
of those structures (based on uncertainty quantification) that should be subjected to reference calculations. 
Hence, computational costs are significantly reduced by such 
selective local refinement of the network data~\cite{Proppe2017,FaradayUncertainty2017,Simm2018,Proppe2019}.

If properly set up by tailored meta-algorithms that control efficient workflows, the autonomous exploration and design of catalytic processes based on reaction networks can be made routinely applicable. Its advantages, compared to standard manual
exploration with standard quantum chemical techniques, are that orders of magnitude more reaction steps can be inspected,
which is key for predictive work that must not miss out on important reaction steps. Obviously, no guarantee of completeness
can be given, but there is no alternative other than autonomous procedures if huge sections of a reaction network shall
be mapped, rather than focusing on a few steps that were considered relevant for some reason (\textit{e.g.}, based on prior experimental
knowledge). 

While this already holds true for a given set of reactants, catalytic processes should be described in open-ended
and rolling reaction network explorations because minute amounts of impurities may interfere in a decisive way. This requires
an interactive option for adding new reactants at any time of an autonomous exploration process, which can then benefit from human
insight that can be exploited as a steering element in the exploration process.

To conclude, autonomous reaction network exploration presents a bright avenue for future computational catalysis as the depth of 
understanding acquired through the wealth of data are unprecedented and increases the probability of unexpected
discoveries made \textit{in silico}.

\section*{Acknowledgments}
\label{sec:acknowledgments}
This publication was created as part of NCCR Catalysis, a National Centre of Competence in Research funded by the Swiss National Science Foundation, and the Swiss Government Excellence Scholarship for Foreign Scholars and Artists.

\section*{Appendix}
\subsection*{Computational Methodology}
\label{sec:computational_methodology}

All data management, quantum chemical calculations, and structure manipulations were conducted within
our general software framework \texttt{SCINE}~\cite{Scine2021}.
Its module \texttt{Chemoton}~\cite{Simm2017,Chemoton2021} 
finds new elementary steps with single ended searches of geometrically aligned structures based on reaction coordinates.
The reaction coordinates are based on reactive sites and directions of attack.
The sites are determined by first-principles-based descriptors or a combinatorial geometric criterion in an exhaustive search as applied in this work.
The directions of attack are derived from least steric hindrance.
Our algorithm then extracts a potential transition state (TS) structure from a given reaction coordinate by pushing together (or pulling apart) two predefined lists of reactive sites with a constant force given as an input parameter.
The force parameter controls the length of individual steps in the trajectory.
The push or pull is stopped when a stop criterion, such as colliding nuclei or a change in bonding, has been reached.
Upon pushing together (or pulling apart) the reactive centers, all atoms besides the reactive sites are continuously relaxed.
This approach allows us to start screenings for potential elementary steps from anywhere on the PES, not necessarily starting at a minimum, the single force parameter does not control the allowed energy barriers, but rather allows to balance the computational costs and efficiency of finding a suitable TS guess because it solely controls the step length.
Smaller step lengths allow for a more accurate location of a potential TS, but require more energy calculations.
The potential TS structure is then refined with an optimization algorithm~\cite{Banerjee1985,Baker1986,Bofill1994,Readuct2020}
and then automatically verified by \textit{intrinsic reaction coordinate} (IRC) optimizations~\cite{Fukui1970}.

The elementary steps between structures are categorized into reactions, which connect compounds.
A compound consists of multiple structures, which share an identical connectivity graph.
The graphs are constructed by our library \texttt{Molassembler}~\cite{Molassembler2020} which provides the functionality for generating graphs and guess structures of conformers based on distance geometry for both organic and inorganic structures~\cite{Sobez2020}.

All calculations were performed by external programs, which can be controlled by
the \texttt{SCINE} interface~\cite{Core2020,Utils2020} that allows to freely select and substitute the underlying physical model.
The available methods range from system-focused parametrization~\cite{Brunken2020}, fast semi-empirical methods~\cite{Sparrow2020,Bannwarth2021}, DFT~\cite{Unsleber2018,Neese2018,Balasubramani2020}, up to highly accurate multi reference calculations~\cite{Baiardi2020}, possibly applying multiscale models~\cite{Muehlbach2018,Brunken2021}.

The uncatalyzed reference network of this work was explored with GFN2 as implemented in the xTB program~\cite{Bannwarth2021,XtbCommit2021}.
Molecular oxygen was calculated in its triplet state.
For all bimolecular combinations of molecules within the exploration, the spin multiplicity was chosen as the sum of the individual multiplicities minus one.
After one or more products were found, the smallest possible multiplicity, \textit{i.e.}, singlet states for molecules with an even number of electrons and doublet states otherwise, was assumed.
Throughout this study electronic energies without zero-point vibrational corrections are considered.
During the exploration, the Hessian was calculated for all newly found structures to confirm them as true minima before making them available for further elementary step trials.

All DFT calculations were carried out with the Perdew--Burke--Ernzerhof (PBE) exchange-correlation functional~\cite{Perdew1996} with D3 dispersion correction~\cite{Grimme2010} and Becke--Johnson damping~\cite{Grimme2011}.
The calculations of the organometallic catalyst were carried out with TURBOMOLE 7.4.1~\cite{Balasubramani2020} with the def2-SVP basis set~\cite{Weigend2005} and density-fitting resolution of the identity through the def2/J auxiliary basis set~\cite{Weigend2006}.
The periodic DFT calculations in the Gaussian Plane Wave (GPW) formalism~\cite{Lippert1997} were carried out with CP2K 8.1~\cite{Kuhne2020} with the MOLOPT-DZVP basis set~\cite{VandeVondele2007} and GTH pseudopotential~\cite{Goedecker1996}, for which we implemented an
interface in \texttt{SCINE}.

The crystal structure of Cu$_2$O was retrieved from the Materialsproject database~\cite{Jain2013} and the (001) surface was generated with pymatgen~\cite{Ong2013,Tran2016}.
The calculations were carried out on a 2$\times$2$\times$3 supercell of the
surface slab consisting of 72 atoms with 15~\AA~vacuum added in the $z$ direction to avoid unphysical interactions of images in this direction.

\subsection*{Conformational clustering}
\label{sec:clustering}
\begin{figure}[H]
	\begin{center}
	   \includegraphics[width=\linewidth]{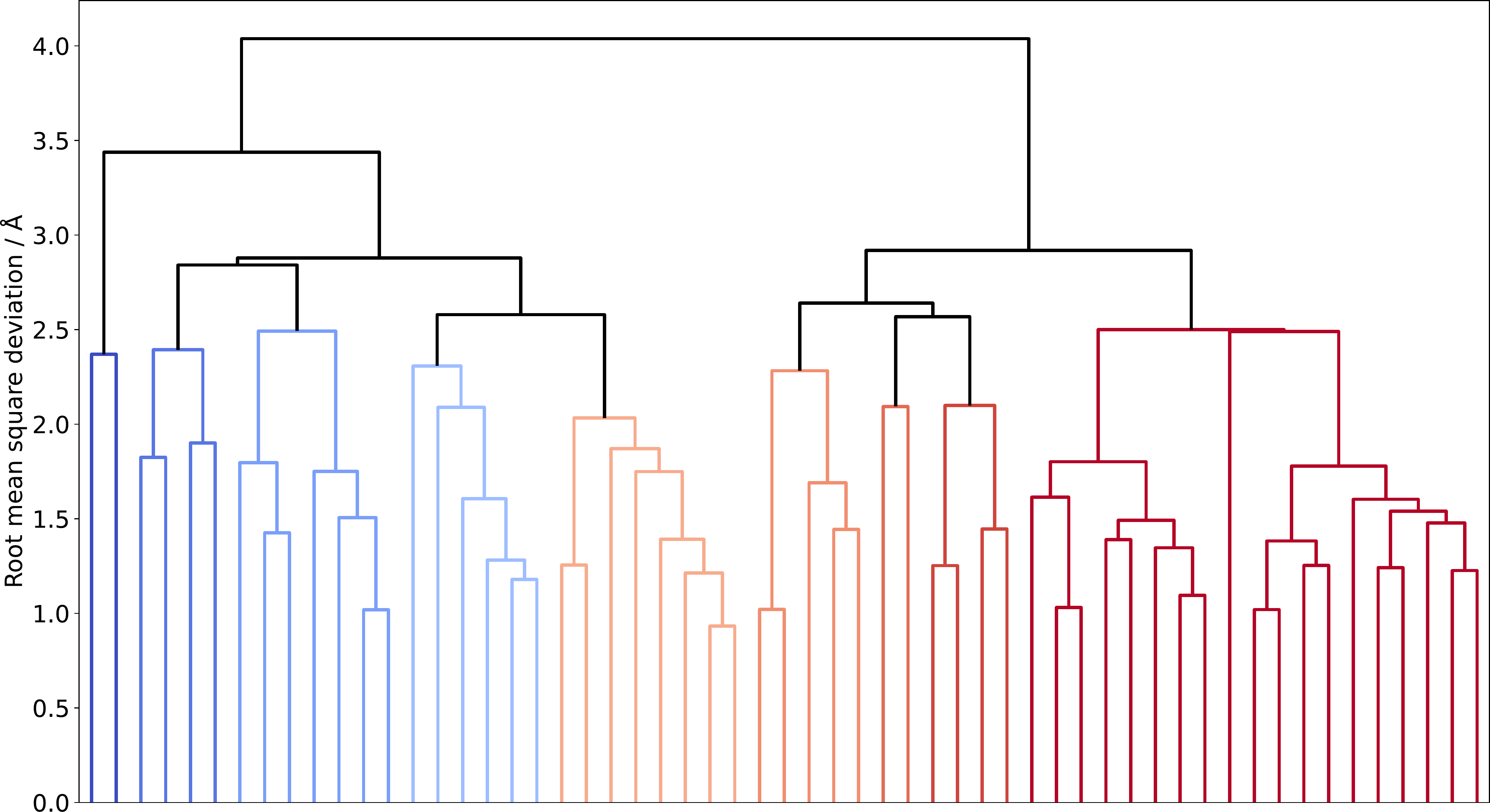}
   \end{center}
   \caption{\label{fig:dendrogram}\small A dendrogram of all 57 optimized conformers generated with \textit{average linkage agglomerative hierarchical} clustering based on the RMSD. Clusters resulting from a cutoff value of 2.5~\AA~are colored.}
\end{figure}

All 57 conformer structures were optimized as outlined above.
The RMSD was calculated for every pair after an optimal alignment.
We then constructed the dendrogram depicted in Fig.~\ref{fig:dendrogram} based on \textit{average linkage agglomerative hierarchical} clustering.
The cutoff value was chosen to be 2.5~\AA~based on inspection of the dendrogram and the resulting centroids of the clusters, which were determined as the structures with the smallest sum of RMSDs to all other structures within the cluster.
The nine representative structures shown in Fig.~\ref{fig:conformers}~B) were, however, not the centroids, but those with the lowest electronic energy.

\providecommand{\latin}[1]{#1}
\makeatletter
\providecommand{\doi}
  {\begingroup\let\do\@makeother\dospecials
  \catcode`\{=1 \catcode`\}=2 \doi@aux}
\providecommand{\doi@aux}[1]{\endgroup\texttt{#1}}
\makeatother
\providecommand*\mcitethebibliography{\thebibliography}
\csname @ifundefined\endcsname{endmcitethebibliography}
  {\let\endmcitethebibliography\endthebibliography}{}

\end{document}